\documentclass{article}

\usepackage{arxiv}

\usepackage[utf8]{inputenc} 
\usepackage[T1]{fontenc}    
\usepackage{hyperref}       
\usepackage{url}            
\usepackage{booktabs}       
\usepackage{amsfonts}       
\usepackage{nicefrac}       
\usepackage[final,nopatch=footnote]{microtype}      
\usepackage{lipsum}		
\usepackage{graphicx}
\usepackage{natbib}
\usepackage{float}
\usepackage{mathtools}
\usepackage{amsmath}
\usepackage{amssymb}
\usepackage{setspace}
\usepackage{hyperref,soul} 
\usepackage{threeparttable}
\usepackage{bbm}
\newcommand{\bbeta}{\boldsymbol{\beta}}
\newcommand{\balpha}{\boldsymbol{\alpha}}
\newcommand{\btheta}{\boldsymbol{\theta}}
\newcommand{\bdelta}{\boldsymbol{\delta}}
\newcommand{\bGamma}{\boldsymbol{\Gamma}}

\newcommand{\bu}{\boldsymbol{u}}
\newcommand{\bx}{\boldsymbol{x}}
\newcommand{\by}{\boldsymbol{y}}
\newcommand{\bz}{\boldsymbol{z}}

\title{Bayesian spatio-temporal modelling for infectious disease outbreak detection}

\author{Matthew Adeoye\\
Department of Statistics\\
University of Warwick
\And Xavier Didelot\\
School of Life Sciences and\\
Department of Statistics\\
University of Warwick 
\And Simon E.F. Spencer\\
Department of Statistics\\
University of Warwick}

\renewcommand{\headeright}{}
\renewcommand{\undertitle}{}
\renewcommand{\shorttitle}{Bayesian spatio-temporal modelling for infectious disease outbreak detection}

\hypersetup{
pdftitle={A template for the arxiv style},
pdfsubject={q-bio.NC, q-bio.QM},
pdfauthor={David S.~Hippocampus, Elias D.~Striatum},
pdfkeywords={First keyword, Second keyword, More},
}

\begin{document}
\maketitle

\begin{abstract}
The Bayesian analysis of infectious disease surveillance data from multiple locations typically involves building and fitting a spatio-temporal model of how the disease spreads in the structured population. Here we present new generally applicable methodology to perform this task. 
We introduce a parsimonious representation of seasonality and a biologically informed specification of the outbreak component to avoid parameter identifiability issues.
We develop a computationally efficient Bayesian inference methodology for the proposed models, including techniques to detect outbreaks by computing marginal posterior probabilities at each spatial location and time point. 
We show that it is possible to efficiently integrate out the discrete parameters associated with outbreak states, 
enabling the use of dynamic Hamiltonian Monte Carlo (HMC) as a complementary alternative to a hybrid Markov chain Monte Carlo (MCMC) algorithm. Furthermore, we introduce a robust Bayesian model comparison framework based on importance sampling to approximate model evidence in high-dimensional space. 
The performance of our methodology is validated through systematic simulation studies, where simulated outbreaks were successfully detected, and our model comparison strategy demonstrates strong reliability. We also apply our new methodology to monthly incidence data on invasive meningococcal disease from 28 European countries. The results highlight outbreaks across multiple countries and months, with model comparison analysis showing that the new specification outperforms previous approaches. The accompanying software is freely available as a R package at \href{https://github.com/Matthewadeoye/DetectOutbreaks}{https://github.com/Matthewadeoye/DetectOutbreaks}.
\end{abstract}


\section{Introduction}

The spread of an infectious disease over space and time is influenced by a large number of interconnected factors 
which can broadly be divided into two categories: factors associated with the host population and factors associated with the infectious disease \citep{anderson1991infectious}. The first category includes how different locations are connected, how populous they are, how immune they are due to either immunisation programmes or past infections, migrations, birth and deaths of individuals. The second category includes properties of the disease such as infectiousness, its natural history, seasonality, mode of transmission, heterogeneities and stochastic effects. To fully account for all these factors is never possible, and therefore a model needs to be considered which simplifies reality while still capturing all the important features \citep{keeling2008modeling,held2019handbook,bjornstad2022epidemics}.
Spatio-temporal modelling has emerged as a fundamental tool for understanding complex systems indexed by space and time. Its interdisciplinary nature and wide-ranging applications in various fields, including ecology, epidemiology, and computer science, have solidified its popularity among researchers and practitioners. In epidemiological research, spatio-temporal models play a critical role in capturing the complexities of disease transmission, accounting for spatial and temporal dependencies in infection dynamics across diverse geographical locations and time periods \citep{hohle2016infectious}.

Bayesian methods have become increasingly prominent in spatio-temporal modelling due to their flexibility in accommodating complex data structures and uncertainty quantification \citep{sahu2022bayesian}. Significant advancements in statistical software packages have facilitated the extensive application Bayesian modelling among practitioners and the development of more sophisticated models by researchers. These advances include the development of notable tools for conducting Bayesian inference such as WinBUGS \citep{lunn2000winbugs}, JAGS \citep{plummer2003jags} and Stan \citep{carpenter2017stan}. 
For example, WinBUGS was used in an analysis of spatio-temporal count data of sheep pox from the Evros region in Greece \citep{malesios2017bayesian}.
However, these programs require a non-trivial level of programming from users. To facilitate accessibility, user-friendly software packages were designed specifically for Bayesian spatio-temporal models. For example, spTimer \citep{bakar2015sptimer} uses Markov chain Monte Carlo (MCMC) techniques to perform Bayesian inference of the parameters of a spatio-temporal model given space-time monitoring data.
Another example is CARBayesST \citep{lee2018spatio} which can use either simple Gaussian random walk Metropolis algorithm or Metropolis adjusted Langevin algorithm (MALA) \citep{roberts1998optimal}.

Despite these advancements, to the best of our knowledge, more complex models such as the one examined in this study cannot be directly implemented using existing software packages. This is due to the latent process of the discrete space-time parameter and the structure of the target posterior density. Performing Bayesian inference for spatio-temporal models remains computationally demanding, and existing approaches struggle to optimize computational efficiency, particularly when incorporating complex outbreak indicators and dynamic temporal patterns. 
Here we introduce a spatio-temporal epidemic modelling framework that builds upon previous research \citep{knorr2003hierarchical, spencer2011detection} while contributing key improvements to the model components. We develop an efficient methodology for Bayesian inference, outbreak detection, and robust model comparison. Through systematic simulation studies, we validate the proposed methods and evaluate the reliability of the model comparison technique. Finally, we demonstrate the practical utility of our framework by applying it to data on the incidence of invasive meningococcal disease across 28 European countries.

\section{Methods}

\subsection{Model structure}
Suppose $i=1,...,I$ represents spatial locations and $t=1,...,T$ represents the time when observed data were recorded. Then $y_{it}$ and $e_{it}$ represent the number of cases and the size of the population at risk at location $i$ at time $t$, respectively. The observed case counts, $y_{it}$, conditional on the logarithmic risks $\lambda_{it}$, are assumed to follow a Poisson distribution as follows:
\[y_{it} | \lambda_{it} \sim \text{Poisson}(e_{it}\lambda_{it}).\] 

Following in the footsteps of previous studies \citep{knorr2003hierarchical, spencer2011detection}, our proposed model involves decomposing the log-risk of disease incidence into a latent trend ($r_t$) component, a spatial component ($u_i$), a modified seasonal component ($s_{t\bmod C}$) and a spatio-temporal outbreak term as follows: 
\begin{equation} \label{eq:model}  
\log(\lambda_{it})=r_t + s_{t\bmod C} + u_i + x_{it}\bz_{it}^T\bbeta.
\end{equation}

\subsubsection{Temporal trend component}
To estimate a smooth background temporal trend, the trend component, $r_t$, is assumed to follow a second-order Gaussian random walk prior given by 
\begin{equation} \label{eq:trendcomp}
r_t | r_{t-1}, r_{t-2} = 2r_{t-1} - r_{t-2} + \epsilon_t, 
\end{equation}
for $t = 3, ..., T$, where $\epsilon_t \sim \mathcal{N}(0, \kappa_r^{-1})$. Uniform priors are assumed for $r_1$ and $r_2$, and this allows the temporal trend component to act as an intercept in the model.

The joint prior density of the trend components $\boldsymbol{r}$ is written as
\[
    \mathbb{P}(\boldsymbol{r}|\kappa_r) \propto \kappa_r^{\frac{T-2}{2}} \exp\left(-\frac{\kappa_r}{2}\sum_{t = 3}^{T}(r_t - 2r_{t-1} + r_{t-2})^2\right),
\]
where $\kappa_r$ is the precision parameter. An alternative formulation of this density is written as
\begin{equation} \label{eq:trend prior}
\mathbb{P}(\boldsymbol{r}|\kappa_r) \propto \kappa_r^{\frac{T-2}{2}} \exp\left(-\frac{\kappa_r}{2}\boldsymbol{r}^T\boldsymbol{R_r}\boldsymbol{r}\right),
\end{equation}
where $\boldsymbol{R_r}$ is a $T\times T$ structure matrix of the second-order Gaussian random walk (RW2) prior, describing the neighbourhood structure of the components:

\[
\boldsymbol{R_r}=\begin{bmatrix*}[r]
1 & -2 & 1 & 0 & 0 & 0 & \ldots & 0 & 0 & 0\\
-2 & 5 & -4 & 1 & 0 & 0 & \ldots & 0 & 0 & 0\\
1 & -4 & 6 & -4 & 1 & 0 & \ldots & 0 & 0 & 0\\
0 & 1 & -4 & 6 & -4 & 1 & \ldots & 0 & 0 & 0\\
\vdots & \vdots & \vdots & \vdots & \vdots & \vdots & \ddots & \vdots & \vdots & \vdots\\
0 & 0 & 0 & 0 & 0 & 0 & \ldots & 1 & -2 & 1\\
\end{bmatrix*}
\]

\subsubsection{Seasonal component}

We propose a first-order cyclic Gaussian random walk model for the seasonal components, $s_t$. This model is formulated such that the seasonal components are repeated throughout the time period, enabling the seasonal component to capture a static repeating pattern. The number of components in the cycle is chosen based on the structure of the available dataset and the epidemiology of the disease under study. 
\[\boldsymbol{s} = (s_1, s_2, \dots , s_{C}),\]
\[s_c - s_{c-1} \sim \mathcal{N}(0, \kappa_{s}^{-1}),\]
$c = 2, \dots , C$ and $s_1 - s_{C} \sim \mathcal{N}(0, \kappa_{s}^{-1})$, so that $s_1$ and $s_{C}$ are considered neighbouring components. The joint density for this prior is written as 
\begin{equation} \label{eq:seasonal prior}
    \mathbb{P}(\boldsymbol{s}|\kappa_s) \propto ({\kappa_s})^{\frac{C - 1}{2}}\exp\left(-\frac{\kappa_s}{2}\boldsymbol{s}^T\boldsymbol{R_c}\boldsymbol{s}\right),
\end{equation}
where $C$ represents the number of components in the cycle, and $\boldsymbol{R_c}$ represents a $C \times C$ structure matrix for the cyclic first-order random walk model (cRW1). The resulting structure matrix is given by: 

\begin{equation*} \label{eq:R_c}
\boldsymbol{R_c}=\begin{bmatrix*}[r]
2 & -1 & 0 & 0 & 0 & 0 &\ldots & -1\\
-1 & 2 & -1 & 0 & 0 & 0 & \ldots & 0\\
0 & -1 & 2 & -1 & 0 & 0 &\ldots & 0\\
0 & 0 & -1 & 2 & -1 & 0 & \ldots & 0\\
\vdots & \vdots & \vdots & \vdots & \vdots & \ddots & \vdots &\vdots\\
-1 & 0 & 0 & 0 & 0 & \ldots & -1 & 2\\
\end{bmatrix*}
\end{equation*}
The sum-to-zero constraint, $\sum_{c=1}^{C} s_c = 0$, is imposed for identifiability of the seasonal components. We implement $C=12$ for both the simulation and application. Thus, each component representing a month in a year and the cycle is forced to repeat over the entire period of the dataset.

\subsubsection{Spatial component}

For the spatial components, $u_i$, we assume a Gaussian intrinsic autoregression, also known as an intrinsic Gaussian Markov Random Field (IGMRF). This model is commonly used in disease mapping in the situation of low counts \citep{besag1991bayesian}. 
\[u_{i}|u_{-i} \sim \mathcal{N}\left(\sum_{j \in n(i)}\frac{u_{j}}{|n(i)|}, \frac{\sigma^2}{|n(i)|}\right),\]
where $n(i)$ is the set of indices of locations that neighbour location $i\in\{1,\dots,I\}$. The joint density of the spatial components $\boldsymbol{u}$ is written as
\begin{equation} \label{eq:spatial prior}
    \mathbb{P}(\boldsymbol{u}|\kappa_u) \propto \left({\frac{\kappa_u}{2\pi}}\right)^{\frac{I-k}{2}}\exp\left(-\frac{\kappa_u}{2}\boldsymbol{u}^T\boldsymbol{R}\boldsymbol{u}\right)\mathbbm{1} (\boldsymbol{u}^T\boldsymbol{1}=0),
\end{equation}
where $\boldsymbol{R}$ with rank $I-k$, is the structure matrix derived from a given adjacency matrix describing the connectivity of the geographical locations under study. The elements of $\boldsymbol{R}$ are
\[
R_{ij} =  \begin{cases}
|n(i)| &\text{if} \ i = j, \\
-1 &\text{if} \ j \in n(i), \\
0 &\text{otherwise}. \end{cases}
\]

Suppose $\boldsymbol{Q}$ is the precision matrix for an intrinsic Gaussian Markov random field, $\boldsymbol{Q}^{-1} = \boldsymbol{\Sigma}$ is its variance-covariance matrix. The matrix $\boldsymbol{Q}$ is rank deficient due to the sum to zero constraint of the structure matrix, $\boldsymbol{R}$. This implies the non-invertibility of $\boldsymbol{Q}$, thus, $\boldsymbol{\Sigma}$ is undefined. Since IGMRFs are improper distributions, they cannot be generative models for data but can be used as priors for modelling \citep{rue2005gaussian}. Here, we derive the elements of the precision matrix for a first-order IGMRF for spatial components.
\[
\text{Prec}(u_{i})=Q_{ii}=\kappa_u|n(i)|
\]

Assuming all spatial components have mean 0,
\[\mathbb{E}(u_{i}|\boldsymbol{u}_{-i})= \sum_{j \in n(i)}\frac{u_{j}}{|n(i)|}= -\frac{1}{Q_{ii}}\sum_{j \in n(i)}Q_{ij}u_{j}\]

 \[\Rightarrow Q_{ij}=  
\begin{cases}
    -\kappa_u & j\in n(i)\\
    \kappa_u |n(i)| & j=i\\
    0 & \text{otherwise.}
\end{cases}\]
Hence, $\boldsymbol{Q} = \kappa_u \times \boldsymbol{R}$.

\subsubsection{Spatio-temporal outbreak component}
The spatio-temporal term $x_{it}$ represents the outbreak indicator with two possible states, the endemic and the hyper-endemic states. In addition, various functional forms for the coefficient of the spatio-temporal term ($\boldsymbol{z}_{it}^T \boldsymbol{\beta}$) in Equation \ref{eq:model} are introduced.

The outbreak indicator ($x_{it}$) is formulated as a hidden Markov model having discrete latent state space. The outbreak indicator visits state 0 during the endemic period and state 1 during the hyper-endemic period. $\boldsymbol{z}_{it}^T$ is a $(p \times 1)$ function of the series at the previous observation time-point, and $\boldsymbol{\beta}$ is a $(1 \times p)$ vector of regression coefficients.

The transitions of the outbreak indicator between the hidden epidemic states is governed by the unknown transition probability matrix, $\boldsymbol{\Gamma}$,

\[\boldsymbol{\Gamma} =
 \begin{pmatrix}
  1-\gamma_{01} & \gamma_{01}\\ 
  \gamma_{10} & 1-\gamma_{10}
\end{pmatrix}  
.\]

Several specifications of the functional form of the outbreak component, $\boldsymbol{z}_{it}^T\boldsymbol{\beta}$ are provided in Table \ref{tab:modspecs}. Models I-VI are taken from \citep{knorr2003hierarchical}. We introduced model VII, which is not prone to the identifiability challenge present in other models. This challenge arises when the absence of a case in the previous time-point extinguishes the effect of an outbreak on the log-risk in the current time-point. This is particularly important in models I, II, and III where $\bz_{it}^T$ is a binary indicator function.  The same challenge can arise in models IV, V, and VI, especially when modelling rare diseases.

An interesting feature of introducing the spatio-temporal component, $x_{it}$, is the possibility of computing the marginal posterior probability of an outbreak in each region at each time point as shown in Section \ref{sec:outBprob}. 

\begin{table}[H]
    \centering
        \begin{threeparttable}
    \caption{\textbf{Outbreak component specifications}}
     \label{tab:modspecs}
  \begin{tabular}{ c c }
        \hline
 Model & $\bz_{it}^T\bbeta$ \\
 \hline
0 & $0$ \\ 
I & $\mathbbm{1}(y_{i, t-1}>0)\beta_1$ \\ 
II & $\mathbbm{1}\left(y_{i, t-1}>0 \ \text{or} \ y_{j, t-1} > 0 \ \ \text{for at least one} \ j \in n(i)\right)\beta_1$ \\ 
III & $\mathbbm{1}\left(y_{i, t-1}>0)\beta_1 \ + \ \mathbbm{1}(y_{j, t-1} > 0 \ \ \text{for at least one} \ j \in n(i)\right)\beta_2$ \\
IV & $\log(y_{i, t-1} + 1)\beta_1$ \\ 
V & $\log(y_{i, t-1} + \sum_{j \in n(i)} y_{j, t-1} + 1)\beta_1$ \\
VI & $\log(y_{i, t-1} + 1)\beta_1 + (\sum_{j \in n(i)} y_{j, t-1} + 1)\beta_2$ \\
VII & $\beta_1$ \\
 \hline
   \end{tabular}
    \begin{tablenotes}
            \item$\mathbbm{1}$ denotes the indicator function
        \end{tablenotes}
    \end{threeparttable}
\end{table}

\subsection{Inference}
Let the parameters be denoted $\boldsymbol{\theta}=(\boldsymbol{r},\boldsymbol{s},\boldsymbol{u}, \kappa_r, \kappa_s, \kappa_u, \boldsymbol{\beta}, \boldsymbol{\Gamma})$, then the target posterior distribution satisfies
\begin{align*}  
\mathbb{P}(\boldsymbol{\theta}|\boldsymbol{y}_{1:I,1:T}) &\propto \mathbb{P}(\by_{1:I,1:T}|\boldsymbol{\theta})\times \mathbb{P}(\boldsymbol{\theta})\\
&\propto \mathbb{P}(\boldsymbol{\theta})\prod_{i=1}^I \mathbb{P}(\by_{i,1:T}|\btheta),
\end{align*}
using conditional independence between locations given $\bu$. The likelihood terms can be decomposed as
\begin{align}
\label{eq:posterior}
\mathbb{P}(\by_{i,1:T}|\btheta)&= \sum_{\boldsymbol{x}_{i,1:T}\in\{0,1\}^T} \mathbb{P}(\boldsymbol{y}_{i,1:T}, \bx_{i,1:T}|\boldsymbol{\theta}).
\end{align}

The sum in Equation \ref{eq:posterior} contains $2^{T}$ terms and should not be calculated using brute force. Previous work used data imputation within the MCMC to obtain samples of $\bx_{i,1:T}$ and calculate the likelihood \citep{knorr2003hierarchical,spencer2011detection}. The $\bx_{i,1:T}$ were sampled in a Gibbs step using the forward filtering backward sampling algorithm \citep{shephard1994partial}. This algorithm exploits dynamic programming to reduce the computational cost to $\mathcal{O}(T)$. In this work we use only the forward filter to obtain the likelihood terms exactly.

\subsubsection{Likelihood via forward filtering}
Here we describe the forward filtering algorithm to compute the likelihood, recalling that $x_{i,t}$ are dependent over time within each spatial location but conditionally independent between spatial locations. 

Define the state space, $\mathbb{S}=\{0,1\}$ for the Markov chain $\boldsymbol{x}_{i,1:T}$, and let $\boldsymbol{\theta}_i=(\boldsymbol{r},\boldsymbol{s},u_i, \boldsymbol{\beta}, \bGamma)$. We assume that the Markov chain starts from its stationary distribution, $\bdelta$. Given the stochastic matrix, $\bGamma$, the stationary distribution satisfies $\bdelta = \bdelta \times \bGamma$ and hence $\bdelta=(\delta_0,\delta_1)=\frac{1}{\gamma_{01}+\gamma_{10}}(\gamma_{10},\gamma_{01})$.
The forward filtering algorithm involves defining a sequence of vectors $\boldsymbol{\alpha}_{i,t}$ with length $|\mathbb{S}|$, as follows:
\begin{align} \label{eq:forward vector}
 \alpha_{i,t}(x_{i,t}) &=\mathbb{P}(y_{i,1:t},x_{i,t}|\boldsymbol{\theta}_i),   
\end{align}
for each $x_{i,t}\in \mathbb{S}$. 
For the initial timepoint we have:
\begin{align*}
\alpha_{i,1}(x_{i,1}) &=\mathbb{P}(y_{i,1},x_{i,1}|\boldsymbol{\theta}_i) \\
&=\mathbb{P}(y_{i,1}|x_{i,1},\boldsymbol{\theta}_i)\mathbb{P}(x_{i,1}|\boldsymbol{\theta}_i) \\
&=\mathbb{P}(y_{i,1}|x_{i,1},\boldsymbol{\theta}_i)\delta_{x_{i,1}}.
\end{align*}

For $t>1$ these forward vectors satisfy the recursion:
\begin{align*}
\alpha_{i,t}(x_{i,t}) &=\sum_{x_{i,t-1}\in \mathbb{S}}\mathbb{P}(y_{i,1:t},x_{i,t},x_{i,t-1}|\boldsymbol{\theta}_i), \\
&=\sum_{x_{i,t-1}\in \mathbb{S}}
\mathbb{P}(y_{i,t}|x_{i,t},y_{i,1:t-1},x_{i,t-1},\btheta_i)
\mathbb{P}(x_{i,t}|y_{i,1:t-1},x_{i,t-1},\btheta_i)
\mathbb{P}(y_{i,1:t-1},x_{i,t-1}|\btheta_i)\\
 &=\sum_{x_{i,t-1}\in \mathbb{S}}
 \mathbb{P}(y_{i,t}|x_{i,t},\boldsymbol{\theta}_i)
 \mathbb{P}(x_{i,t}|x_{i,t-1},\boldsymbol{\theta}_i)
 \alpha_{i,t-1}
 (x_{i,t-1})
\end{align*}

We can deduce the likelihood terms in Equation \ref{eq:posterior} can be calculate from the sum over $|\mathbb{S}|=2$ terms:
\begin{equation*}
\mathbb{P}(\by_{i,1:T}|\btheta_i) =\sum_{x_{i,T}\in \mathbb{S}}\alpha_{i,T}(x_{i,T}).
\end{equation*}

\subsubsection{Priors}
The selection of priors and hyperpriors for inference is as follows: for each of the transition probabilities, a Beta(2, 2) was chosen to down-weight the prior probability at the extremes. For each of the regression coefficients, a Gamma(2, 2) distribution was selected to prevent label switching issues and ensure that the hyper-endemic periods correspond to periods of increased risk. These priors also facilitate a fair model comparison between models I-VII and model 0, where the parameters $\boldsymbol{\Gamma}$ and $\boldsymbol{\beta}$ do not exist, see \citep{johnson2010use}. As previously described, the trend, seasonal and spatial components have prior densities as specified in Equations \ref{eq:trend prior}, \ref{eq:seasonal prior}, and \ref{eq:spatial prior} respectively. A Gamma(1, 0.0001) hyper-prior for the precision parameter of the trend components $\kappa_r$, a Gamma(1, 0.001) for the precision parameter of the seasonal components $\kappa_s$, and a Gamma(1, 0.01) for the precision parameter of the spatial components $\kappa_u$.

\subsubsection{Monte Carlo sampling from the posterior}
\label{sec:MCMCImp}

The target posterior is a high-dimensional probability density with strongly correlated temporal components. The Hamiltonian Monte Carlo (HMC) is known to have significant success in sampling efficiently from these kinds of probability densities, compared to traditional MCMC sampling methods. This is due to its ability to initiate ambitious moves within the state space whilst maintaining reasonable acceptance rates. We implement the dynamic HMC sampler available via \textit{cmdstanr}, an \textit{R} interface to Stan \citep{carpenter2017stan}. This implementation provides users with the ability to accelerate computations via graphics processing unit (GPU) when available. The total runtime for the application section in this paper was about six (6) hours, with model 0 taking only around 6 minutes to fit. We run 4 chains, each with 5000 iterations (2500 warm-up and 2500 sampling) and the target acceptance statistic $\delta = 0.90$. There were no divergences throughout the sampling phase and we monitored the mixing and convergence of the four chains by ensuring that the improved R-hat statistics $\hat{R}$ were lower than 1.05 \citep{vehtari2021rank}.

In addition, we implement an efficient MCMC update scheme to sample from this high-dimensional posterior in our software, \href{https://github.com/Matthewadeoye/DetectOutbreaks}{DetectOutbreaks}. The main code bottlenecks were written separately in \textit{C++}, then we use the Rcpp package \citep{eddelbuettel2011rcpp} which facilitates the seamless integration of \textit{R} and \textit{C++}. The MCMC update scheme is as follows: Given our prior choice, the full conditionals of the precision parameters were available in closed form, so the precision parameters ($\kappa_r$, $\kappa_s$, and $\kappa_u$) were Gibbs sampled. The transition probabilities ($\gamma_{01}$ and $\gamma_{10}$), and the autoregressive coefficients ($\beta_1$ and $\beta_2$) were updated using Metropolis random walk moves. The latent spatial and seasonal components ($\boldsymbol{u}$ and $\boldsymbol{s}$) were inferred using a similar technique, which updates the unconstrained components jointly via an adaptive random walk Metropolis-Hastings move, with the last component being deterministically updated by exploiting the sum-to-zero constraint which aids identifiability of the components. The components of the latent trend ($\boldsymbol{r}$) were updated in blocks using conditional prior proposals \citep{knorr1999conditional, knorr2002block} in addition to the Metropolis-Hastings adaptive random walk scheme \citep{vihola2011stability, spencer2021accelerating}.

\subsubsection{Posterior probability of an outbreak}
\label{sec:outBprob}
The outbreak indicator in the spatio-temporal component ($x_{i,t}$) represents when and where an outbreak is occurring, where 0 indicates the endemic state and 1 indicates the hyper-endemic state. Hence, the posterior probability of an outbreak 
can be obtained by local decoding using the backward sweep of the forward filtering backward sampling algorithm \citep{zucchini2009hidden}.
We are interested in the conditional probability $\mathbb{P}({x_{i,t}}|\boldsymbol{y}_{i,1:T}, \boldsymbol{\theta}_i)$, for each $t=1,\dots,T$.

We define a sequence of $T$ backward vectors, each with length $|\mathbb{S}|$:
\begin{equation} \label{eq:backward vector}    
\beta_{i,t}(x_{i,t}) =\mathbb{P}(\by_{i,(t+1):T}|x_{i,t},\boldsymbol{\theta}_i).
\end{equation}
In particular we have:
\begin{equation*}
\beta_{i,T}(x_{i,T}) =\mathbb{P}(\Omega|x_{i,T},\boldsymbol{\theta}_i) = 1.
\end{equation*}

For $t=T-1,\dots,1$, the backward vectors satisfy the recursion:
\begin{align*}
\beta_{i,t-1}(x_{i,t-1}) &=\mathbb{P}(\by_{i,t:T}|x_{i,t-1},\boldsymbol{\theta}_i)\\
&= \sum_{x_{i,t}\in \mathbb{S}}
\mathbb{P}(\by_{i,t:T}|x_{i,t},\boldsymbol{\theta}_i)
\mathbb{P}(x_{i,t}|x_{i,t-1},\boldsymbol{\theta}_i)\\
 &=\sum_{x_{i,t}\in \mathbb{S}}
 \mathbb{P}(\by_{i,(t+1):T}|x_{i,t},\boldsymbol{\theta}_i)
 \mathbb{P}(y_{i,t}|x_{i,t},\boldsymbol{\theta}_i)
 \mathbb{P}(x_{i,t}|x_{i,t-1},\boldsymbol{\theta}_i)
  \\
 &=\sum_{x_{i,t}\in \mathbb{S}}
\beta_{i,t}(x_{i,t})
 \mathbb{P}(y_{i,t}|x_{i,t}, \boldsymbol{\theta}_i)
 \mathbb{P}(x_{i,t}|x_{i,t-1},\boldsymbol{\theta}_i).
\end{align*}

With the forward $\balpha_t$ and backward $\bbeta_t$ vectors defined as Equations \ref{eq:forward vector} and \ref{eq:backward vector} we have: 
\begin{align*}
\mathbb{P}({x_{i,t}}|\boldsymbol{y}_{i,1:T},\boldsymbol{\theta}_i)
&=\frac{\mathbb{P}(\by_{i,1:T},x_{i,t} |\boldsymbol{\theta}_i)}
{\mathbb{P}(\by_{i,1:T}|\boldsymbol{\theta}_i)}\\
&=\frac{\mathbb{P}(\by_{i,1:t},x_{i,t} |\boldsymbol{\theta}_i)\mathbb{P}(\by_{i,(t+1):T} |x_{i,t},\by_{i,1:t},\boldsymbol{\theta}_i)}{\mathbb{P}(\by_{i,1:T}|\boldsymbol{\theta}_i)}\\
&=\frac{\alpha_{i,t}(x_{i,t}) \beta_{i,t}(x_{i,t})}{\sum_{x_{i,T}\in \mathbb{S}}\alpha_{i,T}(x_{i,T})}. 
\end{align*}

\subsection{Bayesian model comparison and checking}
Previous studies \citep{knorr2003hierarchical} conduct model comparison via Deviance Information Criterion (DIC) \citep{spiegelhalter2002bayesian}. However, the DIC definition assumes that all parameters in the model are continuous and normally distributed. This is not true for the outbreak indicators in the model; moreover, DIC is prone to produce negative estimates for the effective number of parameters, and it is not defined for singular models \citep{vehtari2017practical}. A robust alternative is the standard Bayesian model comparison technique performed by computing the Bayes factor \citep{kass1995bayes}, which is the ratio of posterior probabilities of competing models. To compute the marginal likelihood, we follow the importance sampling method which uses Monte Carlo simulations to estimate the model evidence \citep{touloupou2018efficient}. The integral
\[\pi(\boldsymbol{Y}) = \int_{\Theta} \pi(\boldsymbol{Y}| \boldsymbol{\theta}) \frac{\pi(\boldsymbol{\theta})}{q(\boldsymbol{\theta})} q(\boldsymbol{\theta}) \, \mathrm{d}\boldsymbol{\theta}\]
is estimated by
\[
    \hat{P_q} = \frac{1}{N} \sum_{j=1}^{N} \pi(\boldsymbol{Y}| \boldsymbol{\theta}^{(j)}) \frac{\pi(\boldsymbol{\theta}^{(j)})}{q(\boldsymbol{\theta}^{(j)})},
\]

where the $\btheta^{(j)}$ are sampled from some importance density $q(\btheta)$. The optimal proposal is the posterior distribution and so a good choice is to fit a multivariate normal distribution to posterior samples obtained from MCMC.
To reduce the variance of the estimator, the density of the importance proposal, $q(\boldsymbol{\theta})$, is chosen to be a heavy-tailed multivariate $t$-distribution (with 3 degrees of freedom) centered around the mean of the posterior samples and with scale matrix given by the covariance matrix of the posterior samples. This proposal density, $q(\boldsymbol{\theta})$, therefore, gives a parametric approximation to the posterior density, $\pi(\boldsymbol{\theta}|\boldsymbol{Y})$.

Bayesian model checking was performed by computing the 95\% posterior predictive intervals of the number of cases per time unit and assessing whether the actual numbers fit within these intervals \citep{gelman1996posterior}.

\section{Application to simulated data}

\subsection{Simulation setup}

We performed a simulation study to compare the performance of the models in detecting outbreaks from simulated spatio-temporal datasets of disease incidence.

We began by independently simulating the individual components of the model, including the trend components, seasonal components, spatial components, and the hidden Markov model. These components were fixed to be the same between models to reduce the stochasticity in the comparisons between models.

Simulating the trend component is fairly straightforward through the specified prior model in Equation \ref{eq:trendcomp}. The first and second terms of the trend  component, with RW2 prior, have improper uniform priors and act as the intercept in the model. We chose the first and second components to be -14 to give similar numbers of cases to previously analyzed data in \citep{knorr2003hierarchical}. By setting the two components to be equal, the direction of the resulting trend is purely governed by the stochasticity in the random walk. A high precision parameter of the trend component was chosen, $\kappa_r = 10000$, in order to obtain a smooth pattern in the trend components without extreme variability. This helps to avoid simulations with unrealistically large numbers of disease case counts or rapid stochastic extinction. The seasonal components were simulated using a Fourier term, see Section 5.4 of \citep{hyndman2018forecasting}:
\[
    \boldsymbol{s}= \text{Amplitude} \times \sin{(2\pi f \boldsymbol{c}),}
\]
where the cycle, $\boldsymbol{c} = (1,2, \dots , 12)$, and the frequency $f$ is 1/12. The amplitude was chosen to be 1.4. The hidden Markov model was simulated by sampling the state space with respect to the transition probabilities. We chose $\gamma_{01} = 0.1$ and $\gamma_{10} = 0.2$, implying the system stays on average twice as long in the endemic state compared with the hyper-endemic state. The spatial components were simulated by sampling from the constrained density of the intrinsic Gaussian Markov random field as outlined in the theory of Gaussian Markov random fields, see Section 3.2 of \citep{rue2005gaussian}. The precision parameter for the spatial component was chosen to be $\kappa_u = 25$. The precision matrix was constructed using $\boldsymbol{Q} = \kappa_u \times \boldsymbol{R}$, where $\boldsymbol{R}$ is the structure matrix obtained from the adjacency structure of the geographical locations.

\begin{figure}[!t]
\centerline{\includegraphics[width=6.5in, height=3.5in, keepaspectratio=true]{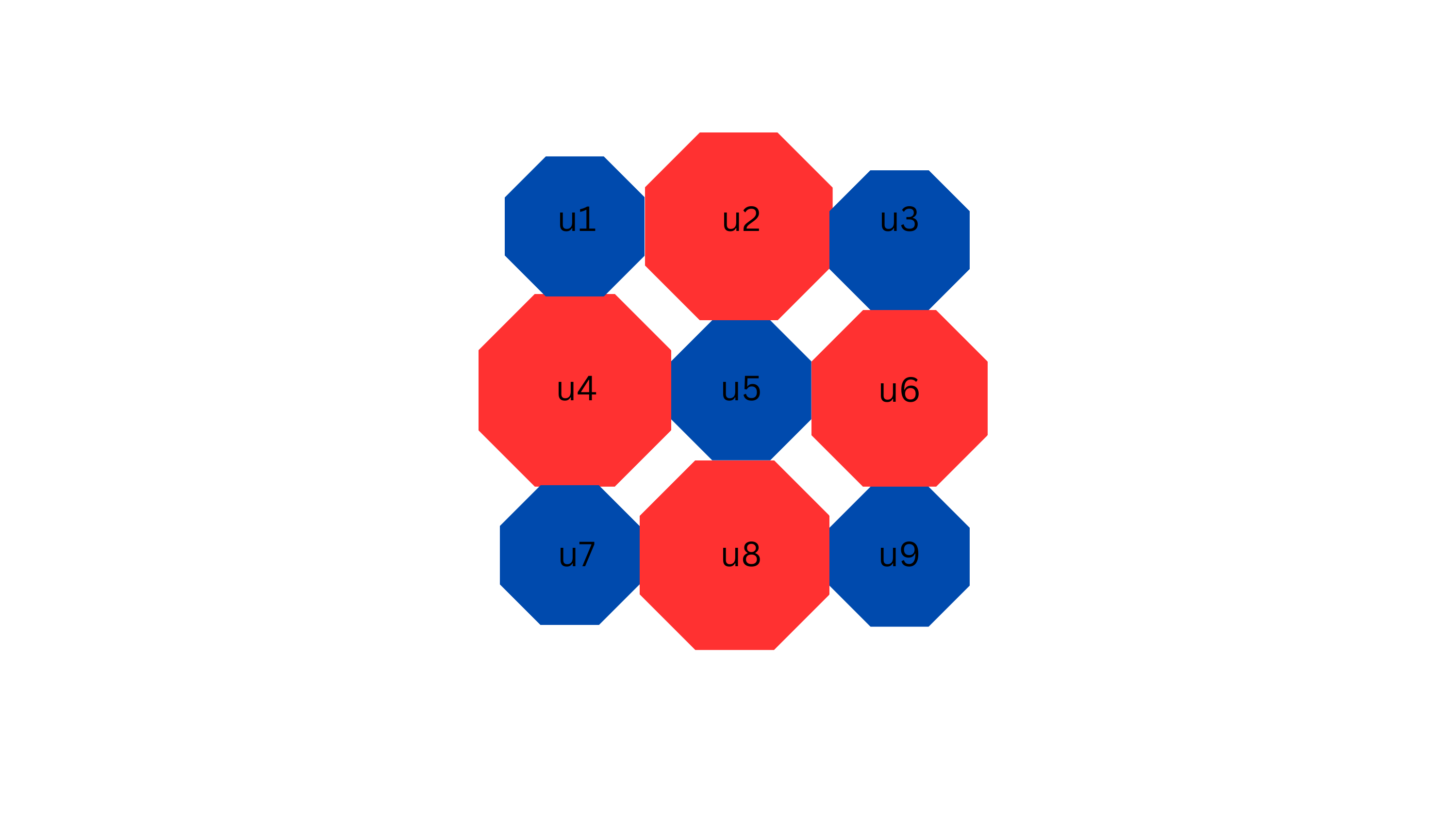}}
\caption{Adjacency structure of spatial locations with small cities in blue and large cities in red.}
\label{fig:simulationCities}
\end{figure}

We assumed there were five (5) small cities each consisting of approximately 500,000 susceptible individuals on average, and four (4) large cities each consisting of approximately one million (1,000,000) susceptible individuals on average. The cities were located such that two cities are considered neighbours if they share a common border. Figure \ref{fig:simulationCities} illustrates the adjacency structure of the hypothetical spatial locations being studied.

For each model specified in Table \ref{tab:modspecs}, we simulate monthly data for a 5-year period in the nine different geographical locations. The regression coefficients in the spatio-temporal outbreak components as shown in Table \ref{tab:simulationParams} were chosen to ensure there is a reasonable contribution to the log risk whenever there is an outbreak. 

\begin{table}[H]
	\caption{\textbf{Parameters for simulation study}}
	\centering
	\begin{tabular}{llllllllll}
		\toprule
		\multicolumn{9}{c}{\textbf{Model}}                   \\
		\cmidrule(r){2-9}
		\textbf{Parameter}  & 0 & I  & II & III & IV & V & VI & VII \\
		\midrule
		$\beta_1$ & -  & 1.65 & 1.65 & 1.25 & 0.55 & 0.49 & 0.35 & 1.65\\
		$\beta_2$     & - & - & - & 0.75 & - & - & 0.20 & -\\
		\bottomrule
	\end{tabular}
	\label{tab:simulationParams}
\end{table}

\subsection{Simulation results}

\begin{figure}[!tp]
\centerline{\includegraphics[width=6.5in, height=3.5in, keepaspectratio=true]{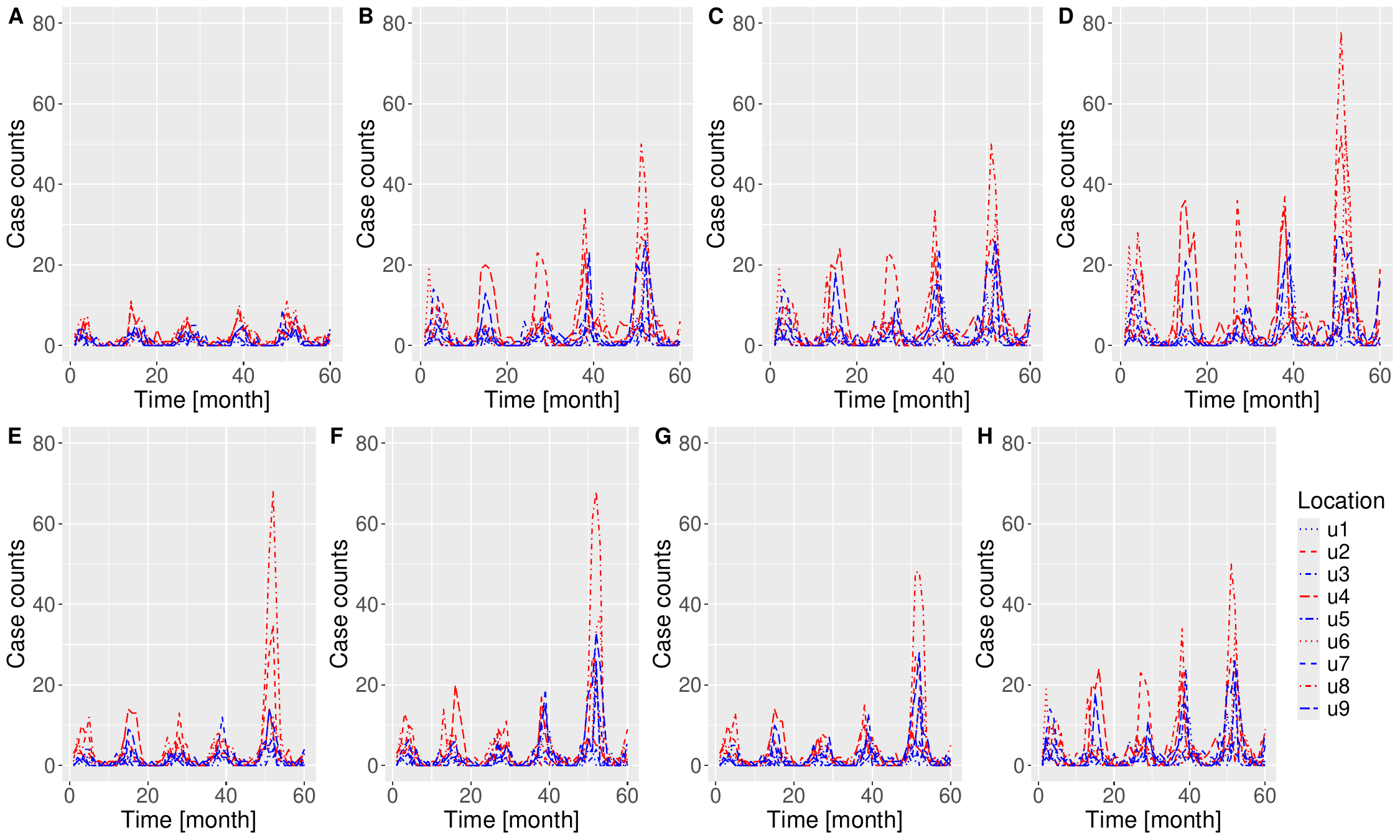}}
\caption{Simulated case counts in each spatial location from (\textbf{A}) model 0, (\textbf{B}) model I, (\textbf{C}) model II, (\textbf{D}) model III, (\textbf{E}) model IV, (\textbf{F}) model V, (\textbf{G}) model VI, and (\textbf{H}) model VII. Small cities shown in blue and large cities in red.}
\label{fig:simulationData}
\end{figure}

Figure \ref{fig:simulationData} shows the number of cases simulated from each of the 8 models. Data from the small cities are shown in blue and the large cities in red. The simulated incidence curves are on a similar scale across the different models, confirming that the parameters in Table \ref{tab:simulationParams} are appropriate. The data for model 0 looks different from the other models with less heterogeneity between regions, as would be expected from the fact that there are no outbreaks in this model. The data for the other models on the other hand show clear evidence for localised outbreaks. Visual inspection of the simulated data in Figure \ref{fig:simulationData} might allow to distinguish model 0 from the others, but seems much less likely to be able to distinguish between the seven other models, which justifies the effort we placed on developing model comparison methodology.

\begin{figure}[!tp]
\centerline{\includegraphics[width=1\textwidth]{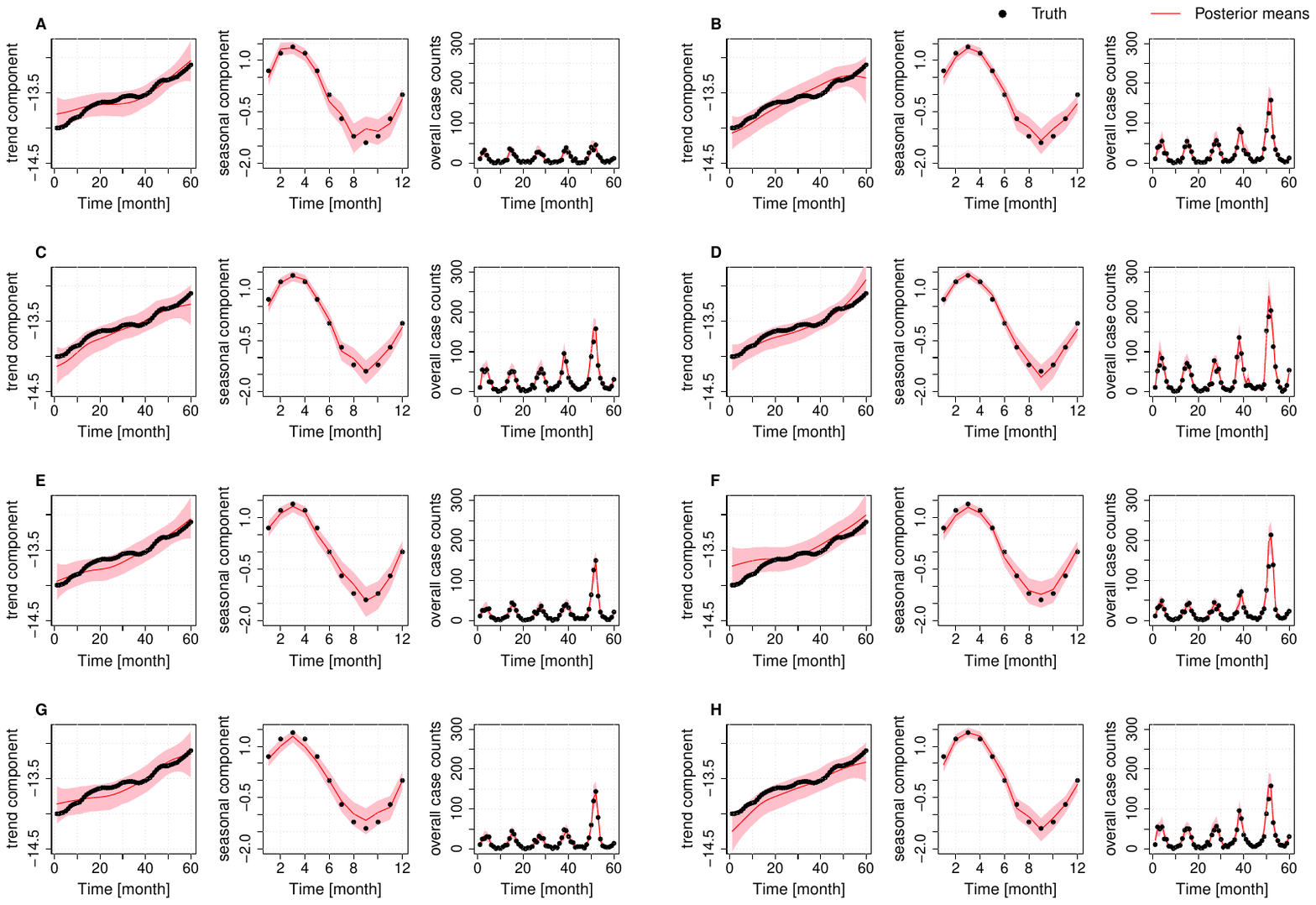}}
\caption{Posterior mean (red line) and credible intervals (shaded pink) for the trend and seasonal components, and the posterior predictive fits for the total case counts across all spatial locations; for (\textbf{A}) model 0, (\textbf{B}) model I, (\textbf{C}) model II, (\textbf{D}) model III, (\textbf{E}) model IV, (\textbf{F}) model V, (\textbf{G}) model VI, and (\textbf{H}) model VII. True values shown with black dots.}
\label{fig:simulationPosteriorpredictive}
\end{figure}

Figure \ref{fig:simulationPosteriorpredictive} shows the posterior means and the associated 95\% credible intervals for the estimated trend and seasonal components across the 8 models, with the true values given by black dots. Both the trend and seasonality component have been inferred well across all models, with the correct values being close to the posterior means and almost always covered by the credible intervals. 
The 95\% posterior predictive credible intervals of the total number of cases per time unit are also shown in Figure \ref{fig:simulationPosteriorpredictive}, and compared with the observed case data. The data is almost always containing within the posterior credible intervals, indicating good model fit to data, as would be expected here since the same model was used for simulation and inference.

Figure \ref{fig:simulationparametersposteriordensities} shows for each of the seven models I-VII the posterior densities of the parameters ruling the outbreak process, namely the transition probabilities $\gamma_{01}$ and $\gamma{10}$, the probability of starting in the outbreak state $\delta_1$ and the autoregressive coefficient $\beta_1$ and $\beta_2$. In each case we find that the posterior mean is close to the true values that were used in the simulations, including the values of the specific parameters given in Table \ref{tab:simulationParams}. 
Figure \ref{fig:simulationspatialcompsposteriordensities} shows the same accuracy in the inferred values of the spatial component parameters.

\begin{figure}[!tp]
\centerline{\includegraphics[width=6.5in, height=3.5in, keepaspectratio=true]{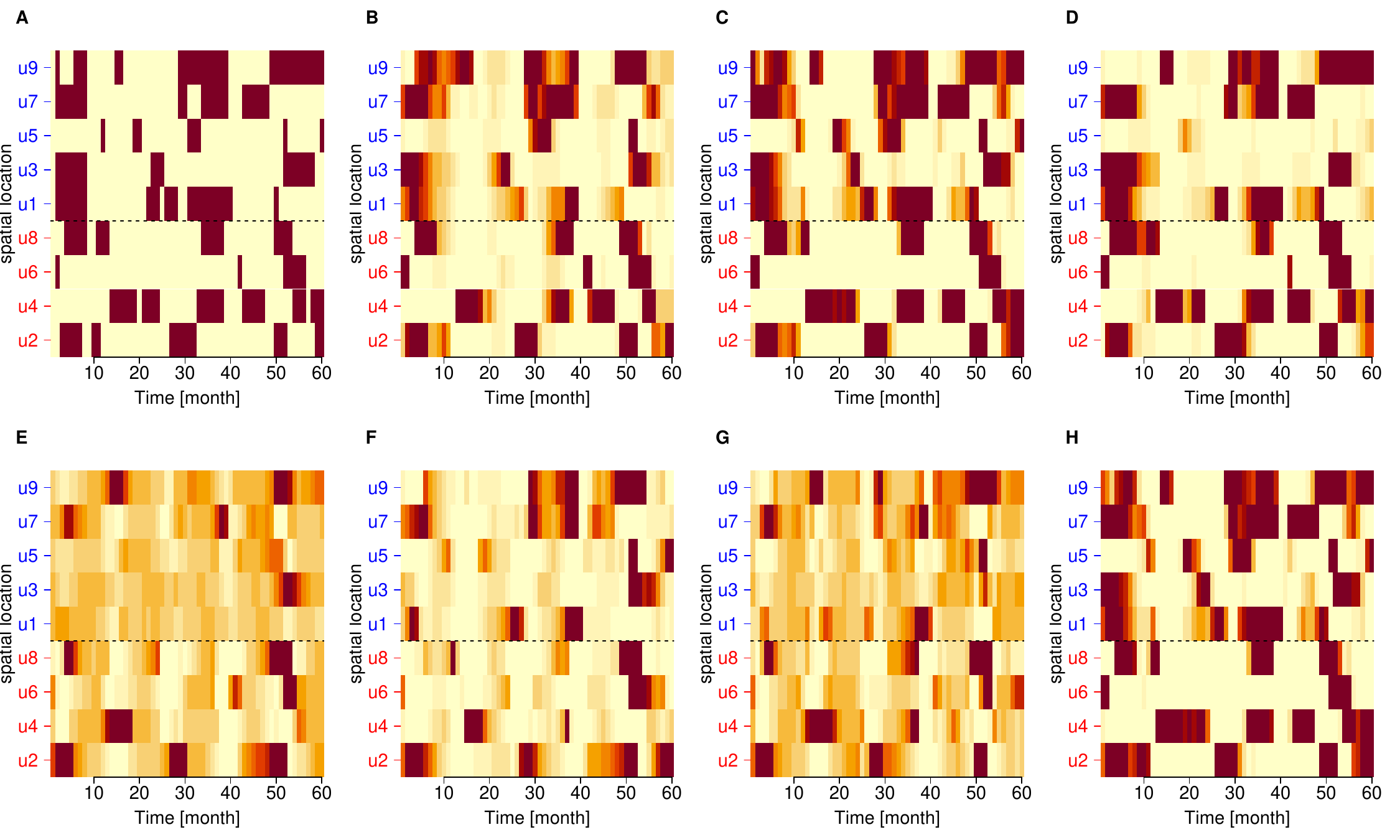}}
\caption{Heat-maps showing (\textbf{A}) Simulated outbreaks (truth) and detected outbreaks using (\textbf{B}) model I, (\textbf{C}) model II, (\textbf{D}) model III, (\textbf{E}) model IV, (\textbf{F}) model V, (\textbf{G}) model VI, and (\textbf{H}) model VII. Small cities placed at the top, and large cities placed at the bottom.}
\label{fig:simulationheatmaps}
\end{figure}

Figure \ref{fig:simulationheatmaps} presents the heatmaps of the simulated outbreaks alongside the outbreaks detected using models I–VII (model 0 is not shown since it does have an outbreak mechanism).
This figure allows us to make qualitative comparisons between the true and inferred occurrence of outbreaks over time and space.
It is visually clear that in all models outbreaks are more likely to be inferred when they truly occurred. However, it is also clear that some models detect outbreaks more precisely. Models IV and VI in particular have less contrasted outbreak detection, with intermediate probabilities between 0.1 and 0.9 more often used than in other outbreaks. In some models, for example model I, it also seems that the outbreaks are more precisely detected for the larger cities, as would be expected from the fact that they would generate more cases and therefore be easier to detect when they occur and to rule out when they do not.

\begin{figure}[!tp]
\centerline{\includegraphics[width=6.5in, height=3.5in, keepaspectratio=true]{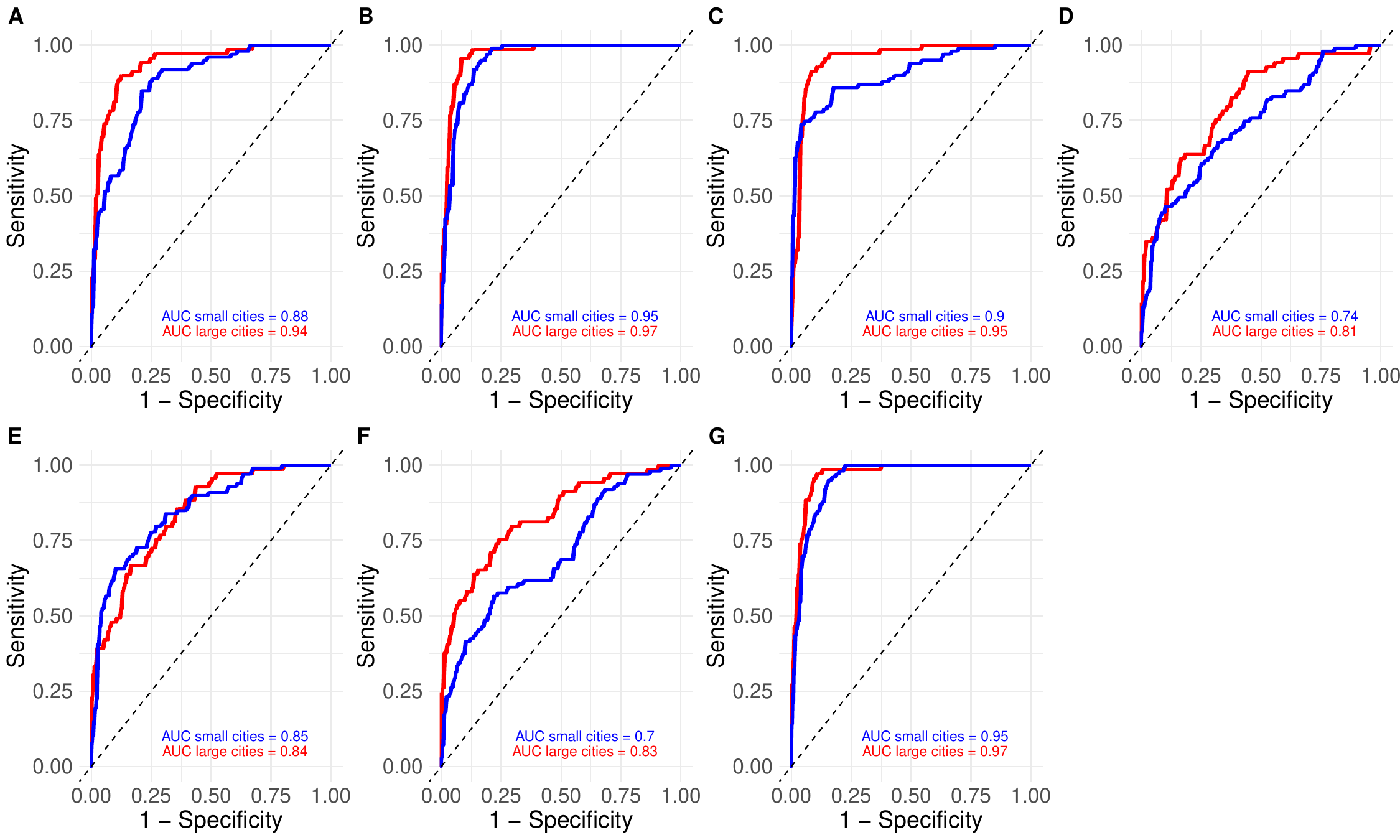}}
\caption{Receiver operating characteristic (ROC) curve for assessing classification accuracy in (\textbf{A}) model I, (\textbf{B}) model II, (\textbf{C}) model III, (\textbf{D}) model IV, (\textbf{E}) model V, (\textbf{F}) model VI, and (\textbf{G}) model VII. Area under the curve (AUC) shown in legend. Small cities shown in blue and large cities shown in red.}
\label{fig:simulationROC}
\end{figure}

To further characterise these results and allow a more quantitative analysis, we present in
Figure \ref{fig:simulationROC} the Receiver Operating Characteristic (ROC) curves for each of these models. This confirms that models I, II, III and VII exhibit superior classification accuracy compared to models IV and VI, with model V being somewhat intermediate. The enhanced performance of models I, II, III, and VII can be attributed to the specification of the spatio-temporal term ($z_{it}$) as outlined in Table \ref{tab:modspecs}, as well as the selected regression parameters for the simulation study in Table \ref{tab:simulationParams}. Specifically, these models preserve the exact effect of the regression coefficient on the log-risks during an outbreak. In contrast, the spatio-temporal term ($z_{it}$) in models IV, V and VI incorporates a log transformation of the number of cases, which diminishes the effect of the regression coefficient on the log-risks (particularly when $z_{it}$ is below 1). Notably, approximately 60\% of the simulated cases from these models are values of zero or one; thus, the outbreak signals in these models are reduced, leading to suboptimal classification performance. Despite their limitations in classification, they can be well-suited for understanding the epidemiological impact of outbreaks on disease risks, particularly in the context of autoregressive effects. Conversely, while models I, II, III and VII provide strong classification capabilities, they can fail to capture nuanced epidemiological effects of prior case magnitudes, as their spatio-temporal term ($z_{it}$) function primarily as binary classifiers of the prior cases. The separation of the heatmaps and ROC curves between large cities and small cities in Figures \ref{fig:simulationheatmaps} and \ref{fig:simulationROC} highlights the strengths of these models in detecting outbreaks within larger populations, especially in models IV and VI where the log transformation in $z_{it}$ now have negligible impact due to the size of disease incidences.

\begin{table}[H]
	    \caption{\textbf{Log marginal likelihoods from simulation study}. The best fitting model is given in bold font.}
	\centering
	\small
	\begin{tabular}{llllllllll}
		\toprule
		\multicolumn{9}{c}{\textbf{Inference model}}                   \\
		\cmidrule(r){2-9}
		\textbf{Simulation model}  & 0 & I  & II & III & IV & V & VI & VII \\
		\midrule
0&\textbf{-687.43}&-688.77&-688.72&-689.99&-688.94&-688.93&-690.13&-687.54\\
I&-1168.05&\textbf{-904.43}&-917.77&-905.45&-977.50&-942.58&-954.29&-916.45\\
II&-1256.41&-998.43&\textbf{-975.23}&-981.31&-1043.28&-1006.16&-1016.34&-978.56\\
III&-1511.78&-953.38&-959.36&\textbf{-949.79}&-1113.21&-1020.67&-1050.05&-966.48\\
IV&-836.88&-773.58&-775.14&-773.37&\textbf{-752.15}&-762.15&-753.08&-774.91\\
V&-1032.70&-885.90&-884.30&-882.49&-902.50&\textbf{-859.87}&-864.81&-885.65\\
VI&-885.50&-819.83&-819.99&-818.69&-805.67&-802.64&\textbf{-799.73}&-815.24\\
VII&-1261.66&-1004.07&-982.16&-987.41&-1046.41&-1010.04&-1020.61&\textbf{-980.94} \\
		\bottomrule
	\end{tabular}
	\label{tab:SystematicSim}
\end{table}

Table \ref{tab:SystematicSim} presents the results of the model comparison analysis based on simulated datasets. Each model was fitted to datasets generated from all models, with the objective of identifying the true data-generating model. The log marginal likelihood estimates demonstrate the effectiveness of this approach in distinguishing between competing models, as in all cases the true data-generating models was the one that had the highest marginal likelihood.
When the data was simulated from model 0, we find that all other models achieve a marginal likelihood only slightly higher than that of model 0. This is because in all models it is possible to effectively switch off the outbreaks, for example by making $\gamma_{01}$ small (so that outbreaks never occur) or by making the $\bbeta$ parameters small (so that when outbreak occur they have little effect). 
When the data was simulated from a model other than model 0, the marginal likelihood of model 0 is always much lower than that of any other model, because the peaks in number of cases caused by outbreaks are difficult to explain in a model without outbreaks, even more so than in a model with an incorrect term for the effect of outbreaks.
Although it is difficult to generalised based on a single dataset from each model, it seems that models I and III are difficult to distinguish from each other, with the same being true for models IV and VI, and for models II and VII, as would be expected from their similar formulations shown in Table \ref{tab:modspecs}.

It is interesting to compare the results of model comparison (Table \ref{tab:SystematicSim}) with model criticism of each model separately applied to each dataset (Figure \ref{fig:syssimulationposteriorpredictivefits}).
The 95\% posterior predictive credible intervals on the total number of infected individuals over time almost always include the actual number of infected individuals not only when the same model was used for simulation and inference (as shown in the diagonal subplots in Figure \ref{fig:syssimulationposteriorpredictivefits}) but also when the models used for simulation and inference differed. 
Even in the most extreme case of model misspecification, which occurs when model 0 is used to fit data from a different model (Table \ref{tab:SystematicSim}), the model criticism is unable to reject the model. This lack of discriminatory power in the model criticism is partly due to the great flexibility of the models we consider here: for example the missing spatio-temporal term in model 0 can be emulated by separate spatial and temporal terms. This result justify the effort we put on designing and formally comparing alternative models.

\section{Application to invasive meningococcal disease}
\begin{figure}[!tp]
\centerline{\includegraphics[width=6.5in, height=3.5in, keepaspectratio=true]{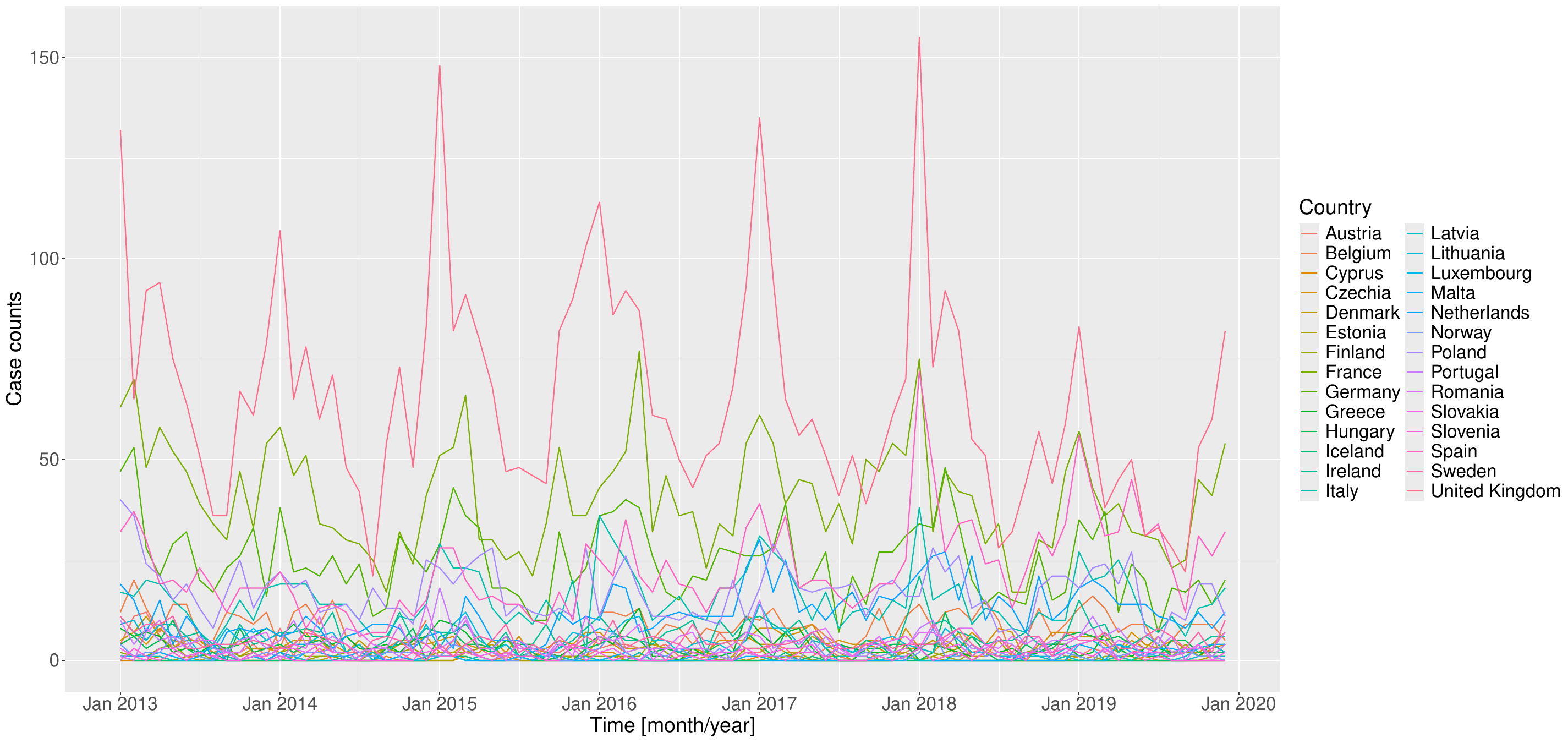}}
\caption{Monthly reported case counts of invasive meningococcal disease in 28 European countries from January 2013 to December 2019.}
\label{fig:applicationdata}
\end{figure}

The European Center for Disease Prevention and Control (ECDC) atlas database contains monthly incidence data on invasive meningococcal disease from 28 European countries \citep{ECDC_Atlas}. The data were available up to December 2021, however, our analysis is restricted to the period from January 2013 to December 2019 to avoid possible confounding factors introduced by the COVID-19 pandemic and Brexit. Although the dataset includes some missing values, our methodology accommodates this under the assumption of data missing at random \citep[MAR; ][]{zucchini2009hidden}. The data is shown in Figure \ref{fig:applicationdata}, from which clear yearly seasonal patterns can be seen. It is also clear that the United Kingdom (UK) has a high number of disease cases, even compared to other countries with a comparably large population such as Germany, France or Italy. This higher incidence may be attributed to increased testing efforts in the UK relative to the other countries. 

Annual population counts for each country were available from the Eurostat database \citep{Eurostat}. To derive approximate monthly population counts, we apply linear interpolation to ensure that the population data align with the temporal resolution of the incidence records.
Geographical data on the distances between the capitals of the countries included in our analysis were obtained from the Cshapes \textit{R} package \citep{gleditsch2010mapping}. Leveraging this data, we classified countries as adjacent if the distance between their capitals was less than or equal to 820 km. This threshold was chosen to ensure that approximately 20\% of pairs of countries were classified as neighbours, providing a balanced representation of geographical proximity in the analysis.

We performed Bayesian inference on this dataset under each of the eight models in Table \ref{tab:modspecs} separately. 
The posterior predictive fits are shown in Figure \ref{fig:applicationposteriorpredictives}. The 95\% posterior predictive credible intervals adequately capture the observed data, indicating a good model fit across all fitted models. However, in model 0, the credible intervals fail to encompass the total case count for January 2018, likely due to the occurrence of an outbreak, which is not accounted for in this model. In contrast, the predictive credible intervals in the other models that incorporate an outbreak component successfully captured this spike.
Other than that all predictive posterior distributions look quite similar, confirming the result we obtained previously on simulated data that this method is not very effective at discriminating between models (Figure \ref{fig:syssimulationposteriorpredictivefits}).

\begin{figure}[!tp]
\centerline{\includegraphics[width=6.5in, height=3.5in, keepaspectratio=true]{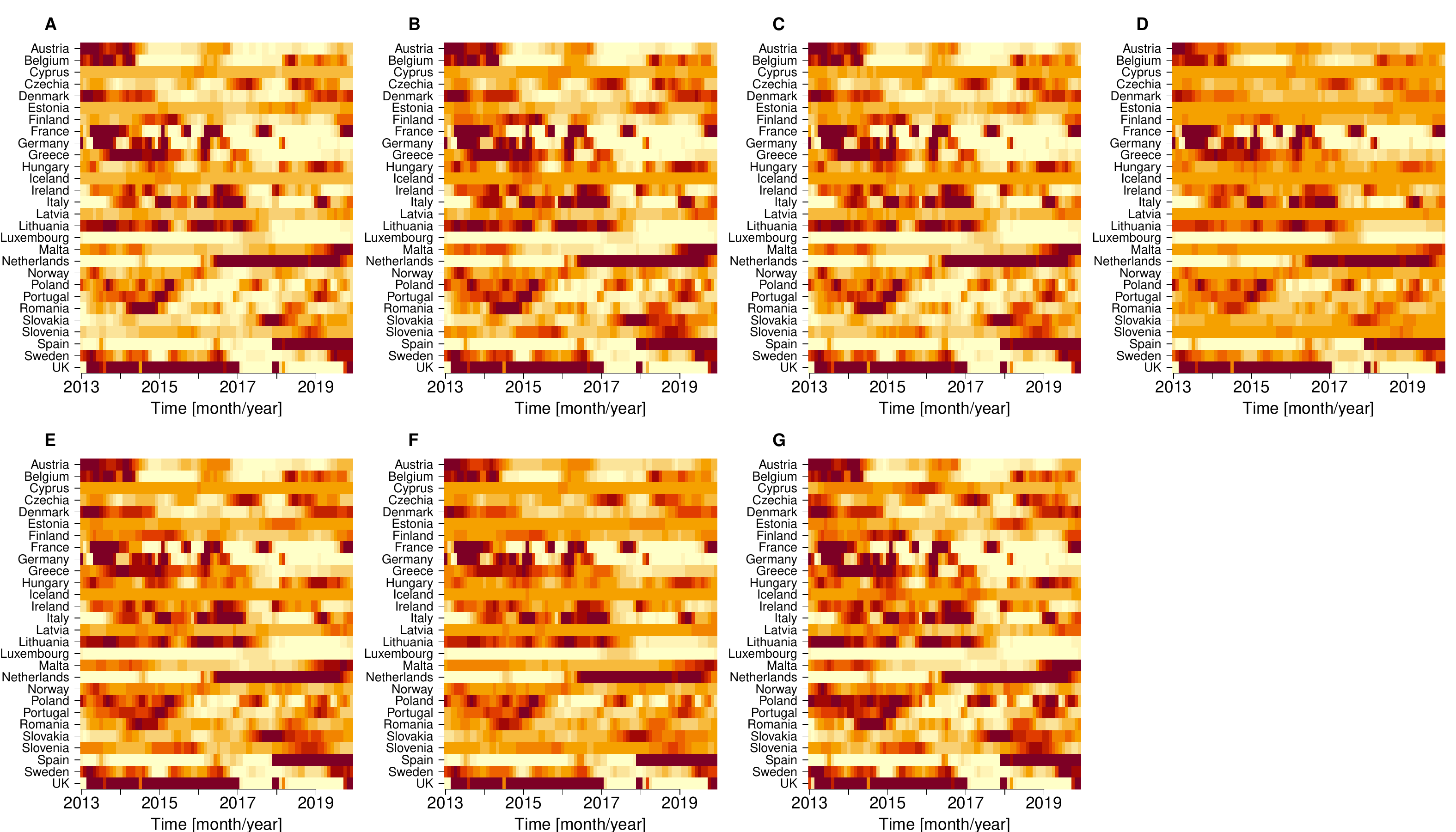}}
\caption{Heat-maps showing for the meningococcal disease application the outbreaks detected using (\textbf{A}) model I, (\textbf{B}) model II, (\textbf{C}) model III, (\textbf{D}) model IV, (\textbf{E}) model V, (\textbf{F}) model VI, and (\textbf{G}) model VII.}
\label{fig:applicationheatmaps}
\end{figure}

Figure \ref{fig:applicationheatmaps} presents the posterior means of the outbreak probabilities across time and countries as captured by models I–VII. There are clear periods of outbreak within each country, with the exception of Luxembourg which shows a low probability of outbreak at all time and across all models. This is likely attributable to the substantial proportion of missing data from this country. Multiple periods of outbreak are observed in France, Germany, and Italy, as well as in several other countries. The correlation is high between the spatio-temporal distributions of outbreaks detected by each model. All correlation coefficients are greater than 0.90, with the lowest one being 0.91 between model IV and model VII (Figure \ref{fig:correlogram}).This means that the detection of outbreaks is relatively robust to the model being used, which is reassuring given that none of the models are likely to perfectly capture the complex reality of how meningitidis spreads within and between countries.

\begin{figure}[!tp]
\centerline{\includegraphics[width=6.5in, height=3.5in, keepaspectratio=true]{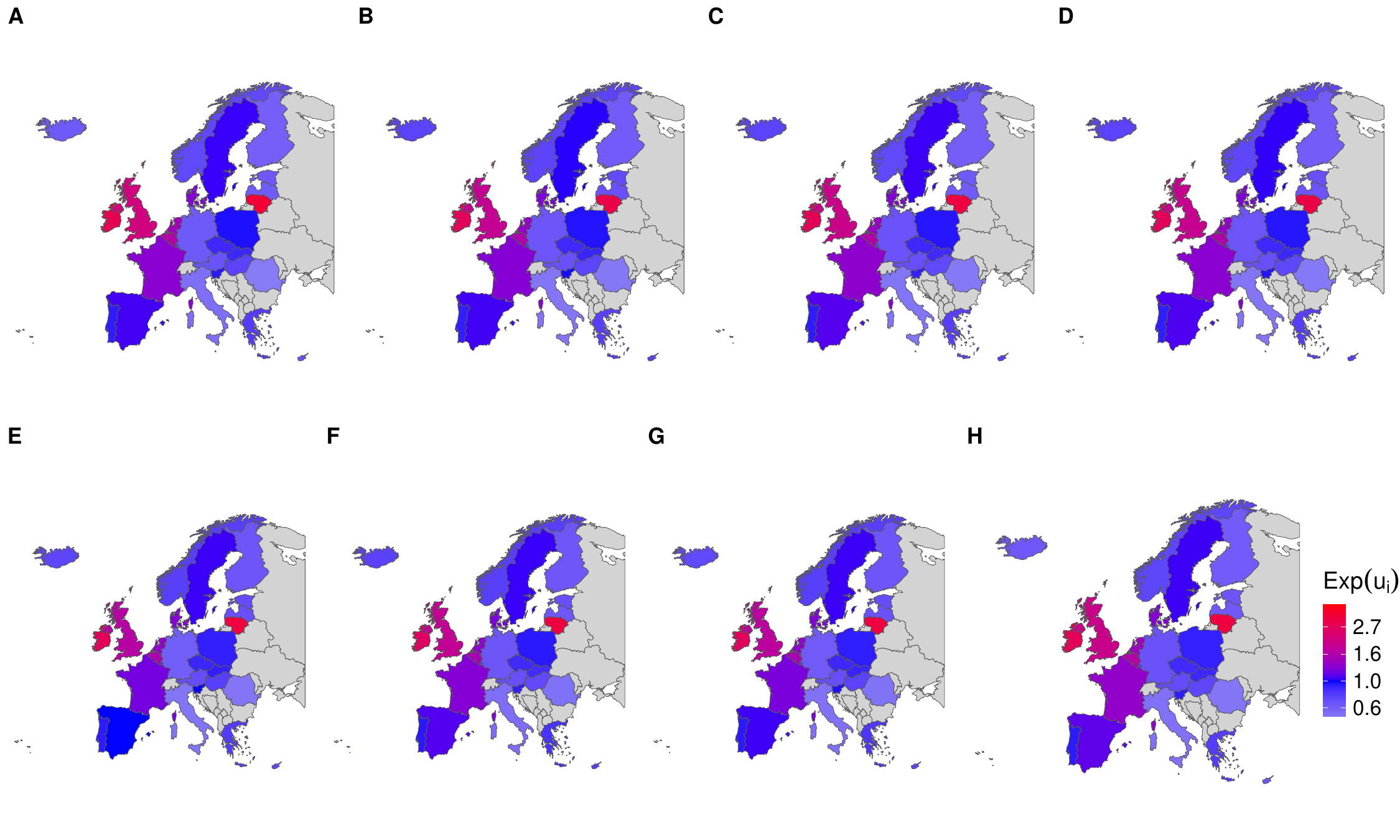}}
\caption{Posterior median relative risks (compared to the geometric mean risk) for the meningococcal disease application from (\textbf{A}) model 0, (\textbf{B}) model I, (\textbf{C}) model II, (\textbf{D}) model III, (\textbf{E}) model IV, (\textbf{F}) model V, (\textbf{G}) model VI, and (\textbf{H}) model VII.}
\label{fig:applicationrelativeriskmap}
\end{figure}

Figure \ref{fig:applicationrelativeriskmap} shows the posterior mean relative risks across Europe as captured by models 0–VII. The relative risks for each location are measured against the geometric mean risk specified in the IGMRF prior (Equation \ref{eq:spatial prior}). The map effectively highlights regions with lower and higher relative risks, with Malta exhibiting the highest relative risk of 4.1. However, due to the small geographic size of Malta, this distinction is not easily observable on the maps. Lithuania, Ireland, and the United Kingdom follow, displaying relative risks higher than all other countries. In contrast, the countries with the lowest relative risks include Romania, Italy, Finland, and Iceland, among others. 
Similar differences between the relative risks across European countries have been described before \citep{nuttens2022evolution}. 
As for the detection of outbreaks (Figure \ref{fig:applicationheatmaps}) we find that the results shown in Figure \ref{fig:applicationrelativeriskmap} for the relative risks in countries are relatively robust to the model choice, even in the case of model 0 which does not have outbreaks. This implies that the differences in risks between countries are real underlying effects and not confounded by the presence of outbreaks or power to detect them.

\begin{figure}[!tp]
\centerline{\includegraphics[width=6.5in, height=3.5in, keepaspectratio=true]{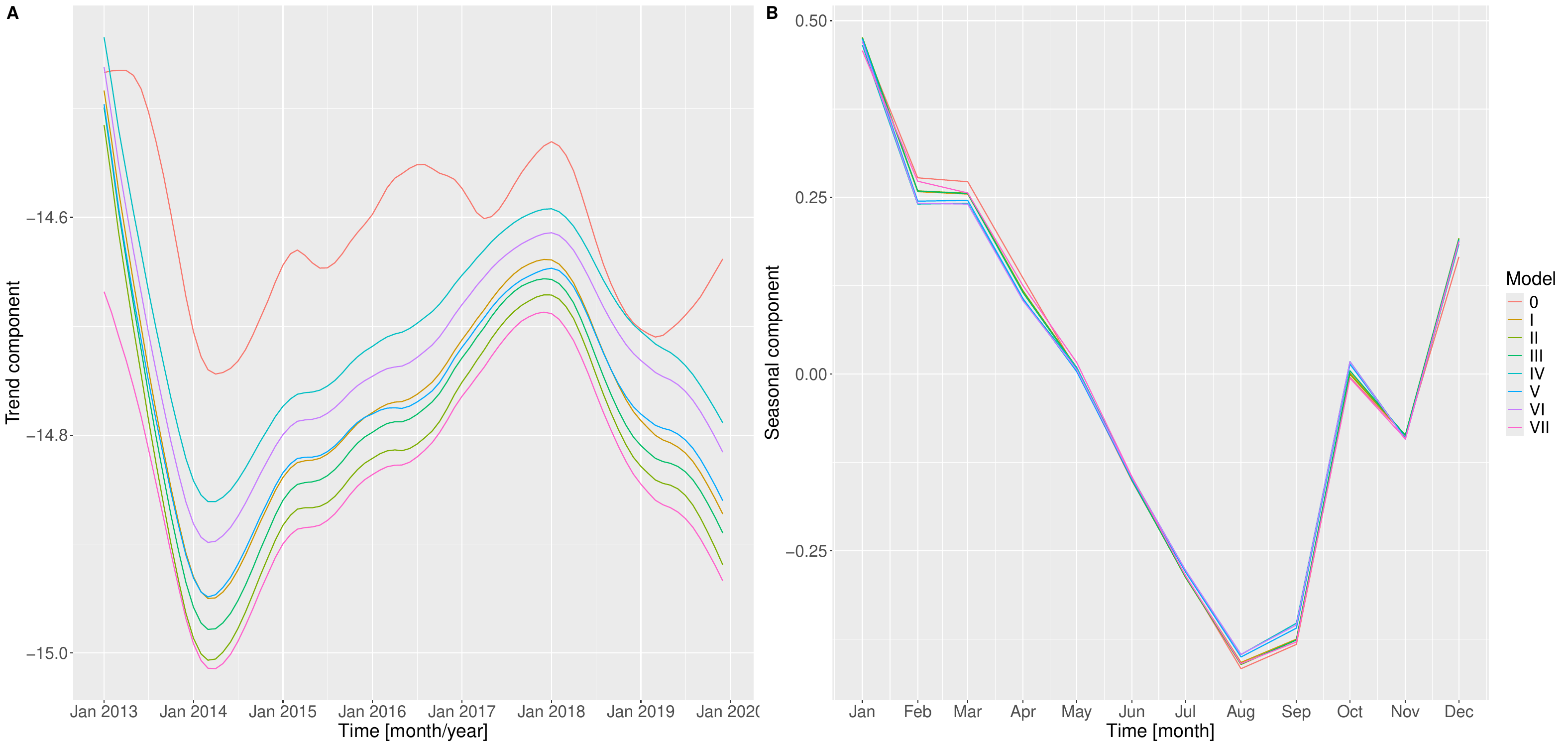}}
\caption{Posterior means for the meningococcal disease application of (\textbf{A}) the trend components, and (\textbf{B}) the seasonal components across all fitted models.}
\label{fig:applicationtrendseasonalcomps}
\end{figure}

Figure \ref{fig:applicationtrendseasonalcomps} illustrates the posterior means of the temporal trend and seasonal components for models 0–VII. The temporal trend component in model 0 is noticeably different from that of models I-VII, with more short term spikes and less long term trends. This suggests that some of short term spikes may be capturing outbreaks in some of the larger countries, since model 0 does not have a separate outbreak mechanism. In contrast, models I–VII incorporate an outbreak component that accounts for a portion of this variability, resulting in a more stable trend component. The trends for models I-VII are very similar, with only a small additive difference between them, but otherwise showing a decrease from 2013 to 2014 followed by a decrease until 2018 and another decrease afterwards. 
The seasonal components are very consistent across all models, identifying January as the month with the highest risk of meningococcal disease, after which a linear decrease in risk operates until reaching the lower point in August, followed by a re-increase until the next January. This seasonal pattern repeats itself for every years as enforced by our prior model (Equation \ref{eq:seasonal prior}). This result is in good agreement with previous analysis of seasonal dynamics of meningicoccal disease \citep{paireau2016seasonal}.

\begin{table}[H]
	\caption{\textbf{Posterior means and 95\% credible intervals in the meningococcal disease application for the autoregressive coefficients, transition probabilities, stationary probability for the hyper-endemic state, and precision parameters.}}
    \tiny
	\centering
	\begin{tabular}{llllllllll}
		\toprule
		\multicolumn{9}{c}{\textbf{Parameter}}                   \\
		\cmidrule(r){2-9}
		\textbf{Model}  & $\beta_1$ & $\beta_2$  & $\gamma_{01}$ & $\gamma_{10}$ & $\delta_1$ & $\kappa_r$ & $\kappa_s$ & $\kappa_u$ \\
		\midrule
		0 & -  & - & - & - & - & 10542 (3171, 22452) & 29.02 (12.79, 51.08) & 0.87 (0.50, 1.33) \\
		I & 0.41 (0.37, 0.44) & - & 0.11 (0.07, 0.16) & 0.16 (0.11, 0.23) & 0.41 (0.32, 0.51) & 16313 (5635, 32570) & 28.36 (12.23, 49.77) & 0.94 (0.54, 1.45)\\
            II & 0.41 (0.38, 0.45) & - & 0.12 (0.08, 0.18) & 0.14 (0.09, 0.21) & 0.46 (0.35, 0.56) & 15284 (5187, 30818) & 28.30 (12.74, 49.29) & 0.92 (0.52, 1.41)\\
            III & 0.24 (0.10, 0.36) & 0.17 (0.05, 0.31) & 0.12 (0.07, 0.17) & 0.15 (0.10, 0.22) & 0.44 (0.33, 0.53) & 15924 (5481, 32080) & 28.07 (12.42, 49.48) & 0.94 (0.53, 1.43)\\
            IV & 0.10 (0.10, 0.11) & - & 0.15 (0.09, 0.21) & 0.15 (0.09, 0.22) & 0.49 (0.39, 0.59) & 19727 (6733, 39213) &  28.73 (12.62, 50.23) & 0.88 (0.50, 1.37)\\
            V & 0.09 (0.08, 0.09) & - & 0.12 (0.07, 0.17) & 0.14 (0.08, 0.20) & 0.46 (0.36, 0.57) & 16745 (5348, 33931) & 28.15 (12.08, 49.87) & 0.94 (0.53, 1.44)\\
            VI & 0.08 (0.05, 0.10) & 0.03 (0.01, 0.05) & 0.13 (0.08, 0.20) & 0.14 (0.09, 0.21) & 0.48 (0.38, 0.58) & 18478 (6188, 37172) & 28.22 (12.61, 49.30) & 0.91 (0.51, 1.41)\\
            VII & 0.42 (0.39, 0.45) & - & 0.13 (0.08, 0.18) & 0.14 (0.09, 0.20) & 0.48 (0.38, 0.57) & 17223 (5376, 36732) & 29.77 (13.08, 52.53) & 0.92 (0.52, 1.42)\\

		\bottomrule
	\end{tabular}
	\label{tab:applicationposteriormeans}
\end{table}

Table \ref{tab:applicationposteriormeans} presents the posterior means and 95\% credible intervals for the model parameters estimated from models 0–VII based on the invasive meningococcal dataset. Notably, the precision parameter of the trend component ($\kappa_r$) in model 0 is considerably lower than in the other models, suggesting a less smooth temporal trend component, as evidenced in Figure \ref{fig:applicationtrendseasonalcomps}A. This can be attributed to the reduced number of parameters in model 0, which allows the trend component to capture more variability in the dataset. In contrast, models I–VII incorporate additional outbreak parameters that help to absorb part of this variability, resulting in a smoother trend component. The precision parameters for the seasonal and spatial components, on  the other hand, are quite lower and exhibit relatively consistent values across all models. This suggests that the seasonal and spatial components exhibit greater heterogeneity compared to the trend component due to the implied higher variances. Specifically, the precision parameter for the spatial components indicates substantial spatial heterogeneity, which may explain the efficiency of their joint updates during the Markov chain Monte Carlo (MCMC) procedure. 

Estimates of the regression coefficient ($\beta_1$) in models I, II and VII suggest an approximately 50\% increase in risks during outbreak periods. A comparable estimate is observed in model III when the regression coefficients ($\beta_1$ and $\beta_2$) are summed, though with wider credible intervals compared to models I, II and VII. In contrast, the estimates from models IV, V and VI indicate small increase in risks, approximately 9\% during outbreak periods with particularly tight credible intervals in models IV and V. The drastic reduction in risks observed in models IV, V, and VI can be attributed to their incorporation of the magnitude of disease incidence from the previous time period, both within a given location and in neighbouring locations (specifically in models V and VI) into the outbreak term ($\boldsymbol{z_{it}}$). In contrast, $\boldsymbol{z_{it}}$ in models I, II, III, and VII binarily classifies the presence or absence of cases, without accounting for the magnitude of these previous incidences. Additionally, the posterior means and credible intervals for the transition probabilities ($\boldsymbol{\Gamma}$) and the stationary distribution of the hyper-endemic state ($\delta_1$) are relatively consistent across all models. The transition probability $\gamma_{01}$ is slightly smaller than $\gamma_{10}$ in most models, indicating that slightly more time is spent in the endemic state rather than the hyper-endemic states, with switches between two states happening on average every 6 to 9 months.

\begin{table}[H]
\centering
        \begin{threeparttable}
    \caption{\textbf{Model comparison for the meningococcal disease application}}
  \begin{tabular}{ c c c}
        \hline
Model & Log marginal likelihood & Posterior model probability\\
 \hline
0&-4998.14& 0.000\\
I&-4850.96& 0.001\\
II&-4846.73& 0.039\\
III&-4850.18& 0.001\\
IV&-4870.95& 0.000\\
V&-4863.87& 0.000\\
VI&-4871.35& 0.000\\
VII&-4843.54& 0.959\\
 \hline
   \end{tabular}
   \label{tab:applicationEvidence}
    \end{threeparttable}
\end{table}

Table \ref{tab:applicationEvidence} presents the results of the model comparison analysis based on the meningococcal dataset. The estimated log marginal likelihood and posterior model probabilities identify model VII as the best-fitting model, with a log marginal likelihood of -4843.54 and a posterior model probability of 0.959. 
The second best-fitting model is model II, although it is found to be almost 16 times less likely than model VII.
This superior performance of model VII can be attributed to its ability to overcome the parameter identifiability issues observed in the other models by allowing only the outbreak indicators to regulate the increase in risk during outbreak periods. 

\section{Discussion}

In this study, we extend a previously described spatio-temporal epidemic modelling framework designed for disease outbreak surveillance \citep{knorr2003hierarchical,spencer2011detection} and propose methodological enhancements to improve its utility and efficiency. In particular, we show how efficient Bayesian inference can be performed by reducing the number of parameters to estimate, including integrating out the outbreak indicators and repeating the seasonal components. Additionally, we present a systematic simulation study and an effective Bayesian model comparison method that leverages importance sampling to approximate the marginal log-likelihood, enabling rigorous Bayesian model selection.
Analysis of simulated data, for which the true model and parameters are known, was conducted to validate the methods presented in this study. By keeping the outbreak indicators fixed across all models, we were able to evaluate the performance of each model in detecting outbreaks as shown by comparing the heatmaps in Figure \ref{fig:simulationheatmaps} and more precisely measured using the ROC curves in Figure \ref{fig:simulationROC}. This modelling framework is well-suited for detecting when and where outbreaks happened in datasets as shown in the simulation study.
A new model VII was introduced to mitigate the identifiability challenge present in the other models: the outbreak indicator is forced to the endemic state whenever its coefficient ($z_{it}\bbeta$) is zero (Table \ref{tab:modspecs}). This happens when there are no cases in the previous timepoint of that spatial location for models I and IV, or including its neighbours for models II, III, V, and VI. 

Our new methodology was applied to seven years worth of epidemiological data on invasive meningococcal disease from 28 European countries \citep{ECDC_Atlas} demonstrating its scalability and practical applicability. The results provide epidemiological insights of public health importance, and illustrate well the capacity of our framework to separately infer heterogeneities caused by spatial (Figure \ref{fig:applicationrelativeriskmap}), temporal (Figure \ref{fig:applicationtrendseasonalcomps}) and spatio-temporal effects (Figure \ref{fig:applicationheatmaps}). The posterior means of the regression coefficients reveal a non-negligible increase in risk when there is an outbreak (Table \ref{tab:applicationposteriormeans}). The associated 95\% credible intervals show uncertainties and statistical evidence of consistent increase in risk during outbreak. There is about 50\% increase in risk during outbreak in model I, II, and VII. Across all models, we see evidence of increased risk during outbreak. 
Despite assuming a uniform prior across all eight models, we found that model VII outperformed the other models in this application with a posterior model probability of 0.959 
(Table \ref{tab:applicationEvidence}). 

The inference algorithms presented in this study offer significant computational efficiency and flexibility for practical applications. By integrating out the outbreak indicators from the likelihood, the number of parameters to be estimated and the Monte Carlo error introduced by sampling the additional parameters is drastically reduced. The prior model introduced for the seasonal component (Equation \ref{eq:seasonal prior}) contributes to the computational efficiency of the method, especially when analyzing very large datasets.  We implemented Hamiltonian Monte Carlo (HMC) via Stan \citep{carpenter2017stan}, as well as an efficient bespoke Markov chain Monte Carlo (MCMC) sampling scheme, offering flexibility based on computational resources and user preferences. This allowed us to compare these two approaches to inference on the same model and dataset, and we found that they both had pros and cons.
The computational demands of the HMC sampler are known to increase with the number of parameters, which can be a concern given that the number of model parameters grows with the dataset's dimensionality. Overall, we found that the HMC sampler performed best when the datasets was relatively small (for example in the simulated datasets we analyzed) whereas we recommend the use of the bespoke MCMC sampler for very large datasets (such as the meningococcal disease dataset we analyzed). We tried to optimise both methods to make the comparison as fair as possible. In the custom MCMC sampler, main code bottlenecks were rewritten in \textit{C++} thanks to the Rcpp package \citep{eddelbuettel2011rcpp}, significantly accelerating computations compared to our first implementation completely written in R. In the HMC sampler, we leveraged Stan's OpenCL to enhance computational efficiency for users with access to GPUs.

In conclusion, our study contributes methodological innovations and practical tools to advance spatio-temporal epidemic modelling, with implications for both research and real-world disease surveillance applications. 
To facilitate adoption by other researchers and practitioners, we provide a user-friendly software package 
\href{https://github.com/Matthewadeoye/DetectOutbreaks}{DetectOutbreaks}
capable of performing inference and comparison for all models discussed. The models and software have been designed to be as generally applicable as possible, but there are still limitations. For example, it cannot effectively capture opposing trends in different geographical locations. This limitation would become particularly relevant when analyzing data from regions with distinct epidemic dynamics, for example if several countries implemented a vaccination programme whereas several others did not. It is not possible of course for a single modelling framework to capture every such complex scenario that can occur in real datasets. However, even in such a situation our framework should provide a useful starting point for analysis: applying an unrealistic model is often the starting point to discover the features of the data that are not represented and therefore inform the development of better models.

\bibliographystyle{elsarticle-harv}
\bibliography{references}  

\newpage
\pagenumbering{gobble}
\setcounter{figure}{0}
\setcounter{table}{0}
\makeatletter 
\renewcommand{\thefigure}{S\@arabic\c@figure} 
\renewcommand{\thetable}{S\@arabic\c@table} 
\makeatother

\section*{Supplementary Material}

\begin{figure}[H]
\centerline{\includegraphics[width=6.5in, height=3.5in, keepaspectratio=true]{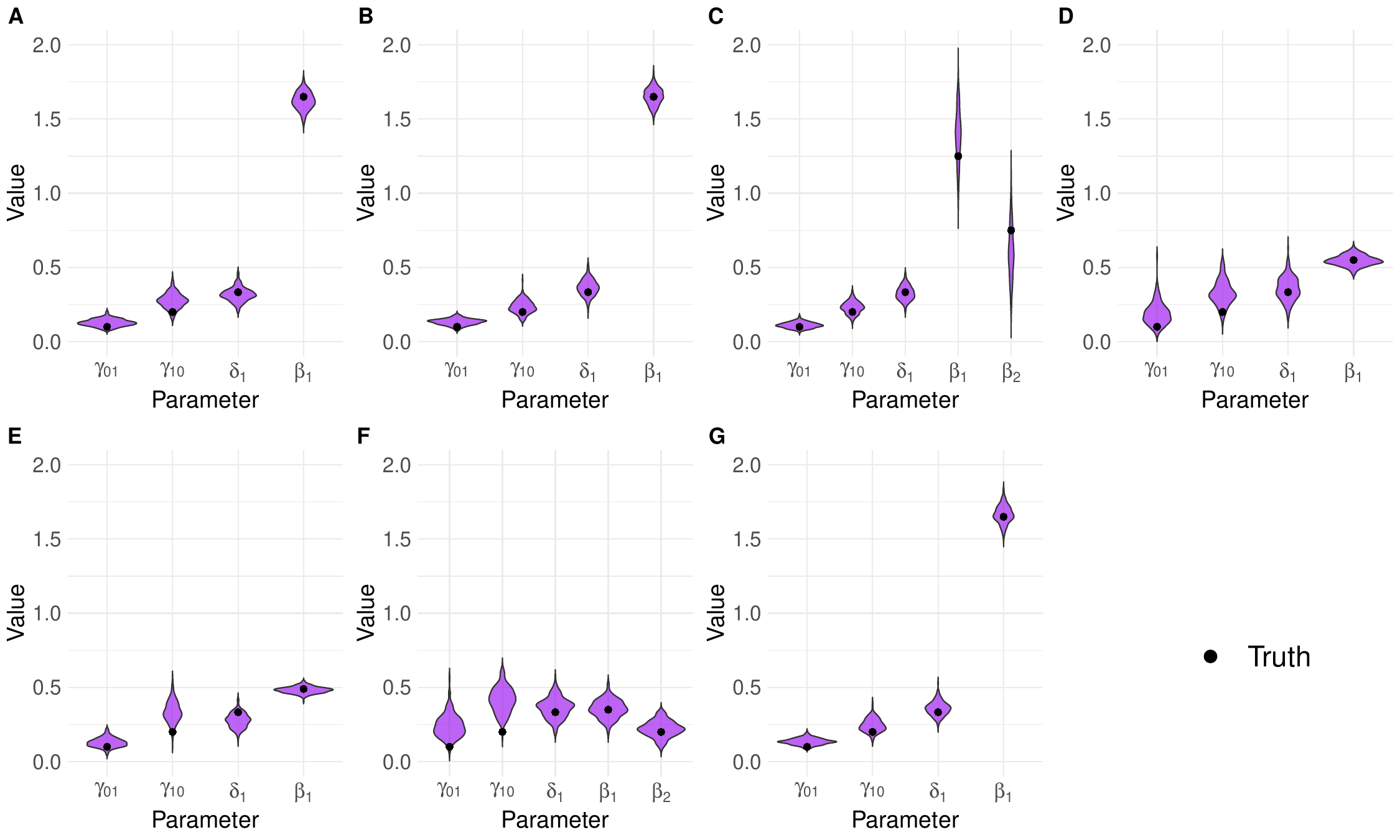}}
\caption{Posterior densities for transition probabilities ($\gamma_{01}$ and $\gamma_{10}$), stationary distribution ($\delta_1$), and autoregressive coefficients ($\beta_1$ and $\beta_2$) for (\textbf{A}) model I, (\textbf{B}) model II, (\textbf{C}) model III, (\textbf{D}) model IV, (\textbf{E}) model V, (\textbf{F}) model VI, and (\textbf{G}) model VII. True values in black dots.}
\label{fig:simulationparametersposteriordensities}
\end{figure}

\begin{figure}[H]
\centerline{\includegraphics[width=6.5in, height=3.5in, keepaspectratio=true]{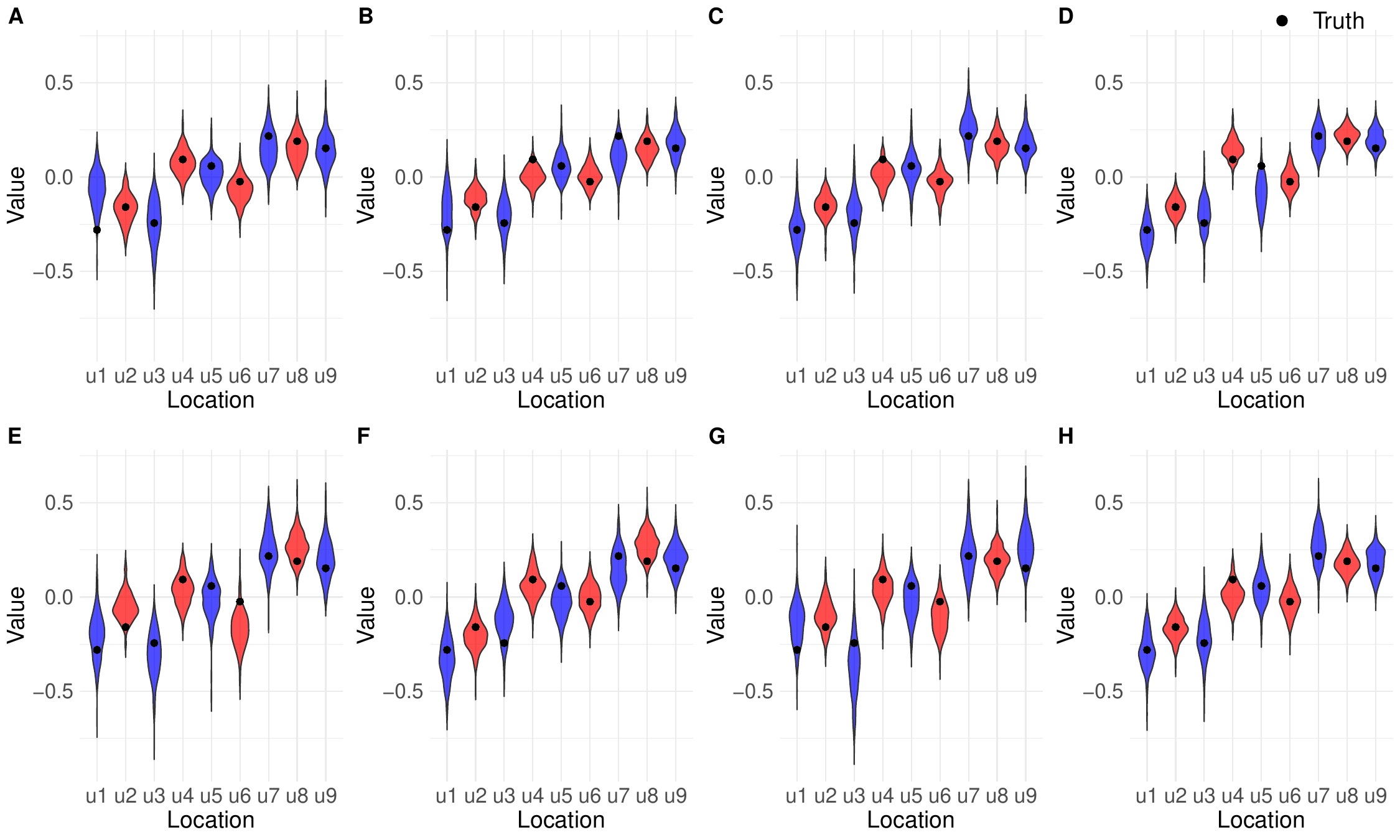}}
\caption{Posterior densities for spatial components with true values in black dots for (\textbf{A}) model 0, (\textbf{B}) model I, (\textbf{C}) model II, (\textbf{D}) model III, (\textbf{E}) model IV, (\textbf{F}) model V, (\textbf{G}) model VI, and (\textbf{H}) model VII.}
\label{fig:simulationspatialcompsposteriordensities}
\end{figure}

\begin{figure}[H]
\centerline{\includegraphics[width=1\textwidth]{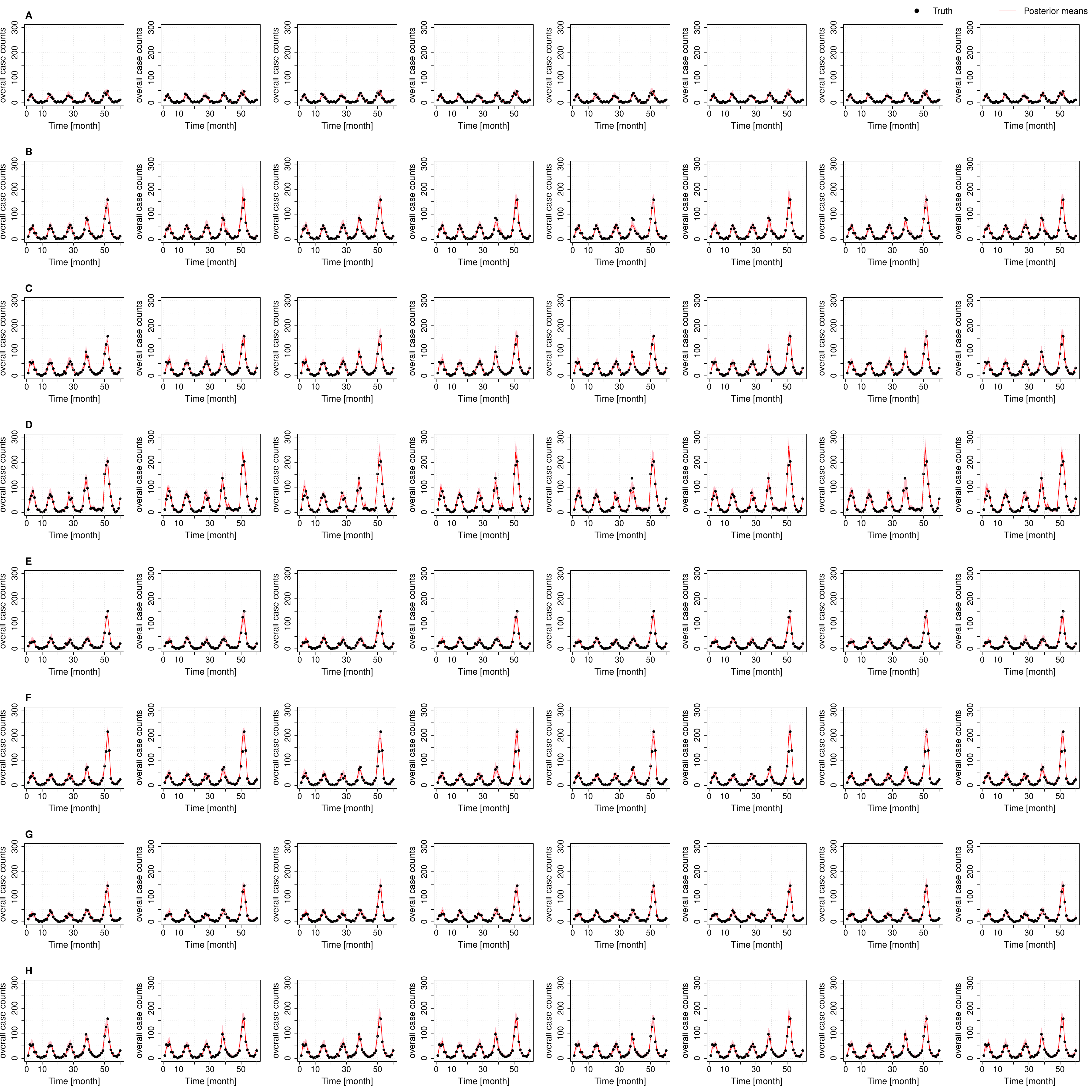}}
\caption{Posterior predictive fits for the total case counts across all spatial locations; for dataset simulated from (\textbf{A}) model 0, (\textbf{B}) model I, (\textbf{C}) model II, (\textbf{D}) model III, (\textbf{E}) model IV, (\textbf{F}) model V, (\textbf{G}) model VI, and (\textbf{H}) model VII, with each fitted to models 0-VII. True values shown with black dots.}
\label{fig:syssimulationposteriorpredictivefits}
\end{figure}

\begin{figure}[H]
\centerline{\includegraphics[width=6.5in, height=3.5in, keepaspectratio=true]{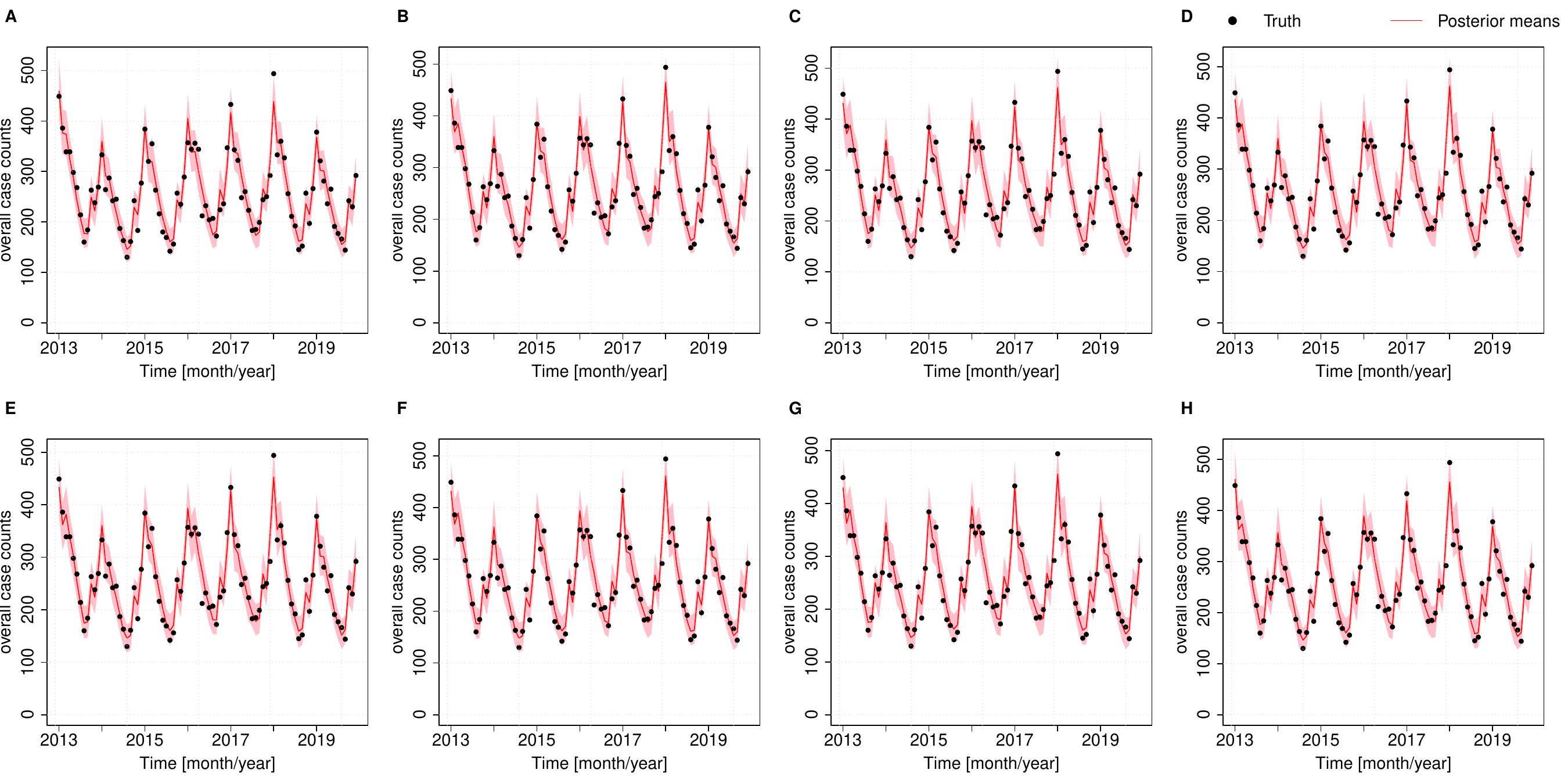}}
\caption{Posterior predictive fits for the meningococcal disease analysis for the overall case counts across all spatial locations for (\textbf{A}) model 0, (\textbf{B}) model I, (\textbf{C}) model II, (\textbf{D}) model III, (\textbf{E}) model IV, (\textbf{F}) model V, (\textbf{G}) model VI, and (\textbf{H}) model VII. Observed data shown with black dots.}
\label{fig:applicationposteriorpredictives}
\end{figure}

\begin{figure}[H]
\centerline{\includegraphics[width=6.5in, height=3.5in, keepaspectratio=true]{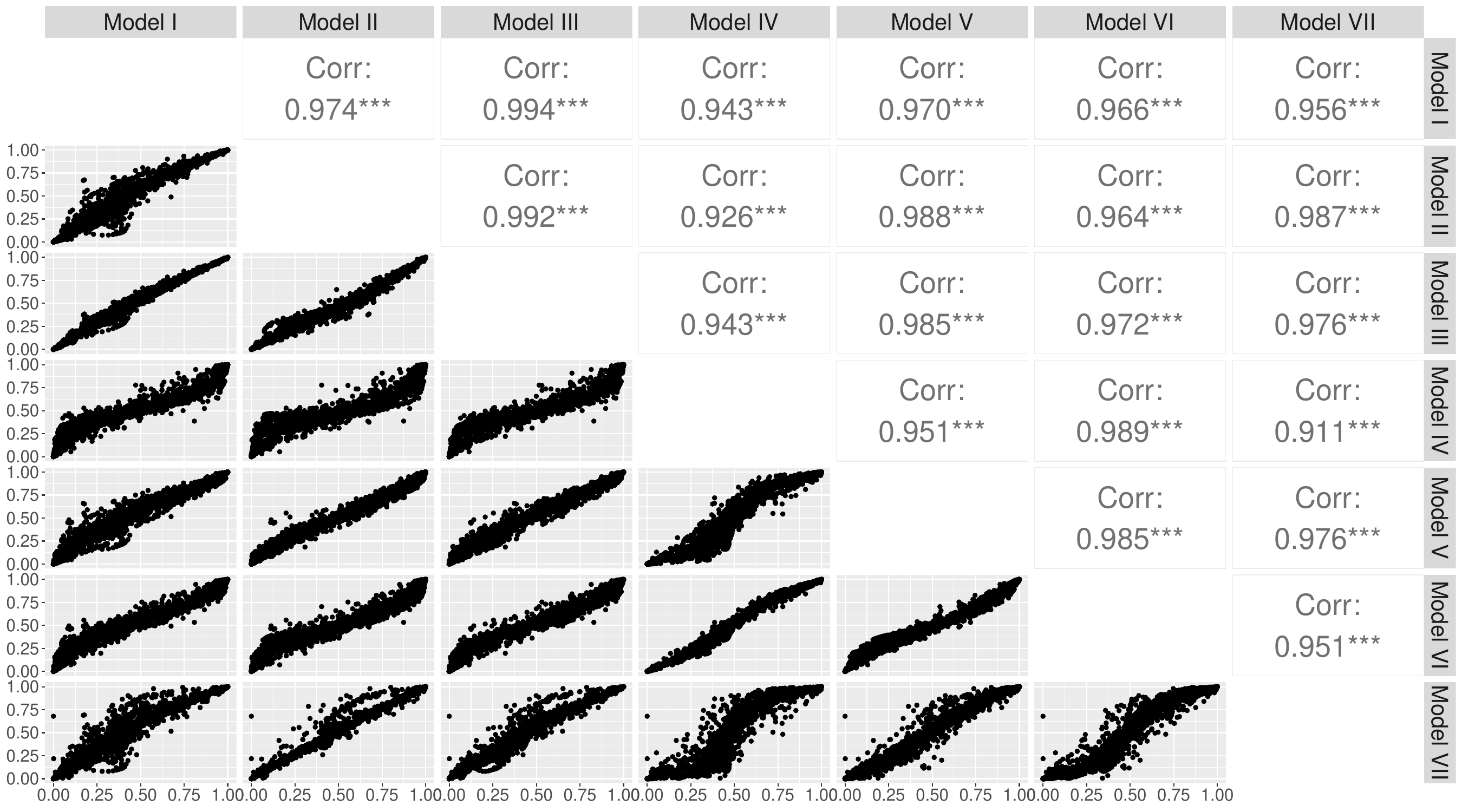}}
\caption{Correlogram for posterior probabilities of outbreak from models I-VII.}
\label{fig:correlogram}
\end{figure}

\end{document}